\documentclass[%
 reprint,
 amsmath,amssymb,
 aps,
]{revtex4-2}

\usepackage{xcolor}
\usepackage{hyperref}

\hypersetup{
    colorlinks,
    linkcolor={red!80!black},
    citecolor={blue!80!black},
    urlcolor={blue!80!black}
}
\usepackage{comment}
\usepackage{units}
\usepackage{amsmath}
\usepackage{booktabs}
\usepackage{graphicx}
\usepackage{dcolumn}
\usepackage{bm}
\usepackage{float}
\usepackage{array} 
\usepackage{booktabs} 
\usepackage{makecell}
\usepackage{soul}
\usepackage [english]{babel}
\usepackage [autostyle, english = american]{csquotes}
\usepackage{multirow}

\UseRawInputEncoding

\MakeOuterQuote{"}

\makeatletter
\renewcommand{\p@subsection}{}
\makeatother

\graphicspath{{./}{figures/}}

\begin{document}

\title{Self-lensing flares from black hole binaries V: systematic searches in LSST}

\author{Kevin Park$^{1,2,3}$}
 \thanks{EMAIL: ksp2136@columbia.edu}
 \author{Zolt\'an Haiman$^{1,4,2,3}$}
 \author{Chengcheng Xin$^{4,5}$}
 \author{Tzuken Shen$^{5}$}
\author{Ashley Villar$^{5}$}
\author{Jordy Davelaar$^{6}$} \thanks{NASA Hubble Fellowship Program, Einstein Fellow}

\affiliation{
$^{1}$Institute of Science and Technology Austria (ISTA), Am Campus 1, Klosterneuburg 3400, Austria}
\affiliation{
$^{2}$ Department of Physics, Columbia University, 550 W 120th St, New York, NY 10027, USA}
\affiliation{
$^{3}$Astrophysics Laboratory, Columbia University, 550 W 120th St, New York, NY 10027, USA}
\affiliation{
$^{4}$Department of Astronomy, Columbia University, New York, NY 10027, USA}
\affiliation{
$^{5}$Center for Astrophysics | Harvard \& Smithsonian, 60 Garden Street, Cambridge, MA 02138, USA}
\affiliation{
$^{6}$Department of Astrophysical Sciences, Peyton Hall, Princeton University, Princeton, NJ 08544, USA
}
\begin{abstract} 
The Vera C. Rubin Observatory has now seen first light, and over a 10 year duration, LSST is projected to catalogue tens of millions of quasars, many of which are expected to be associated with sub-parsec supermassive black hole binaries (SMBHBs). Out of these SMBHBs, up to thousands of relatively massive binary-quasars are expected to exhibit gravitational self-lensing flares (SLFs) that last for at least 20-30 days. We assess the effectiveness of the Lomb-Scargle (LS) periodogram and matched filters (MFs) as methods for systematic searches for these binaries, using toy-models of hydrodynamical, Doppler, and self-lensing variability from equal-mass, eccentric SMBHBs. We inject SLFs into random realizations of damped random walk (DRW) lightcurves, representing stochastic quasar variability, and compute the LS periodogram with and without the SLF. We find that periodograms of SLF+DRW light-curves do not have maximum peak heights that could not arise from DRW-only periodograms.
On the other hand, the matched filter signal-to-noise ratio (SNR) can distinguish SLFs from noise even with LSST-like cadences and DRW noise. Furthermore, we develop a three-step procedure with matched filters, which can also recover injected binary parameters from these light-curves. We expect this method to be computationally efficient enough to be applicable to millions of quasar light-curves in LSST.
\end{abstract}

\maketitle
\section{Introduction}
\label{sec:Intro}

Supermassive black holes (SMBHs) with masses between $10^{6-9} {\rm M}_\odot$  have been observed in the cores of most nearby galaxies \cite{Kormendy2013}. If galaxies primarily grow through mergers, then the two central SMBHs should settle at the center of a merged galaxy driven by dynamical interactions with their environment. Although the fraction of systems that harden to close separations is still an open question, gravitationally bound SMBH binaries (SMBHBs) should be common in galactic nuclei \cite{begelman1980}. Hydrodynamical simulations of the electromagnetic (EM) emission from a SMBHB accreting from circumbinary gas show that this emission can be bright and also periodic long before merger \cite{farris2015, tang2018, Krauth2023, dittmann+2023}.

There are dozens of observations of dual active galactic nuclei (AGN)--two active SMBHs in a single galaxy separated at kpc scales \cite{DeRosa+2019}. However, there are no directly imaged sub-pc SMBHBs due to their close separation.

Currently, one of the most prominent methods to find sub-pc SMBHBs is to find apparent periodicity in quasar/AGN light-curves \cite{d_orazio_2023, bogdanovic2022}. Systematic searches for periodicity in AGN light-curves have been performed for large time-domain optical surveys including the Catalina Real-time Transient Survey (CRTS) \cite{Graham+2015b}, the Zwicky Transient Facility (ZTF) \cite{Chen+2024}, the Palomar Transient Factory (PTF) \cite{Charisi+2016}, and GAIA \cite{huisje_2025}, identifying hundreds of SMBHB candidates (but see also \cite{Jakob_null, Sesana_2018, El-Badry_2025}). 

In addition, recently, pulsar timing arrays (PTAs) have detected a stochastic gravitational wave background (GWB) in the nHz bands, which is consistent with a cosmological population of coalescing SMBHBs \cite{NANOGrav15-GWB, EPTA+InPTA-GWB, Parkes-GWB, ChinesePTA-GWB}. Targeted continuous wave (CW) searches with PTAs have further identified one AGN (dubbed Rohan) which shows modest Bayes-factor support for a CW signal and has sustained quasi-periodicity in its optical light-curve for $\sim 20$ years \cite{agarwal2025nanograv}. Gravitational waves from SMBHBs will also be directly detectable by the Laser Interferometer Space Antenna (LISA)\cite{amaro-seoane2017}, planned for launch in the 2030s. Thus, at the current stage, compiling a catalog of periodic quasars can later enable combining EM and GW signals from the same sources \cite{Xin2024}, which would allow for new tests of general relativity, high energy physics, and cosmology \cite{baker2019}. The Vera C. Rubin Observatory's Legacy Survey of Space and Time (LSST), which has now seen first light in 2025, is expected to contain 20-100 million quasars \cite{Xin_Haiman_2021} over a 10 year survey duration and can contribute significantly to this catalog of periodic AGN. 

Most systematic searches for SMBHB candidates in the time-domain have relied on Lomb-Scargle (LS) periodograms or similar sinusoid-fitting methods, assuming that the periodicity of the emission roughly follows the orbital period of the SMBHB and is sinusoidal. However, it is known that quasars exhibit stochastic red-noise variability which is well described by a "Damped Random Walk" (DRW) model \cite{MacLeod_2010} or a "Damped Harmonic Oscillator" model \cite{yu2025}. This quasar red-noise variability can mimic sinusoid-like periodicities for a few periods \cite{Vaughan+2016, elbadry2025}, meaning many SMBHB candidates, without additional evidence, can exhibit spurious periodicity by chance.

A SMBHB signature that would be more difficult to mimic by red-noise is a "self-lensing flare" (SLF). General relativistic ray-tracing models \cite{davelaar2022, davelaar2022b} (hereafter Papers I, II) show that self-lensing flares occur when binaries are inclined nearly edge-on and the two SMBHs are aligned to the line of sight within an Einstein radius. Then the emission from the "source" SMBH passing behind is gravitationally lensed and magnified by the "lens" SMBH in front, leading to a symmetric flare in the light-curve occuring twice per orbit, assuming both BHs are actively accreting. \cite{Krauth_2024} (hereafter Paper III) shows using hydrodynamical simulations that these flares exist in light-curves even in the presence of circumbinary gas and \cite{park_2025} (hereafter Paper IV) estimate that up to thousands of self-lensing flares at least 20-30 days long, with many more narrower flares, can be expected in LSST. This paper (Paper V) considers the question of how to recover these flares from noisy and sparse light-curves.

There are currently a handful of self-lensing binary candidates reported. An AGN in the Kepler catalog, dubbed Spikey, also had a light-curve which was well-fit with a relativistic Doppler Boost + SLF model from an eccentric binary. The hypothesized self-lensing flare occured as a symmetric $\sim10\%$ magnification at the expected orbital phase--when the Doppler modulation is at a minimum \cite{Hu2020}. Radio observations of Spikey have shown evidence for a wobbling jet \cite{Kun_2020}, which would be consistent with a SMBHB with one BH having an active and precessing jet, but a second flare was not found. Meanwhile, follow-up observations of Spikey with the \textit{Chandra Space Telescope} at a predicted flare (2020 March-May) \cite{sorabella_2022} were inconclusive, as the intrinsic X-ray variability was larger than the amplitude of the self-lensing flare and the relativistic Doppler Boost. 

In addition, \cite{Kollatschny_2024} report a symmetric flare in the AGN NGC 1566, which lasted for roughly 155 days and is well-fit by a microlensing curve with a lens mass of $5\times 10^5 {\rm M}_\odot$. The primary SMBH mass was estimated to be $5\times 10^6 {\rm M}_\odot$ from optical broad emission lines and the SMBHB has an inferred orbital period of roughly 4000 years.

While there are currently only a few uncertain candidates, \cite{Kelley_2021} have also showed that several hundred self-lensing flares lasting 30 days or longer are expected in LSST \cite{LSST_2009} using the Illustris simulation \cite{Sijacki_2015}. This is consistent with Paper IV, which was based on the observed quasar luminosity function (QLF) under the assumption that most AGN are SMBH binaries. Notably, this assumption is supported by the fact that the population of quasars from the QLF yields a gravitational wave background consistent with the recent PTA detections \cite{Kis-Tóth_2025}. Thus, a systematic method tailored to finding periodic flares in large datasets such as those expected soon from LSST AGN is warranted. 

In addition, for a fraction of self-lensing binaries, the angular separation at alignment can be comparable to the size of the source BH's shadow ($\sqrt{27}GM_{\rm bh}/c^2$), which leads to a "dip" in luminosity near the peak of a flare when the BH shadow is lensed. Papers I, II and \cite{porter2024} show that the height and width of the dips can vary depending on the SMBHB parameters and that these dips, along with the self-lensing flares, also exist in light-curves even in the presence of circumbinary gas (Paper III). Paper IV finds that dozens of binaries will be inclined enough to exhibit dips in LSST. Finally, in edge-on ultra-compact binaries, the binary's orbital period will evolve due to the emission of gravitational waves and thus a series of "chirping" self-lensing flares could be observable in an even smaller subset of sources \cite{xin_2025, Haiman2017}.

In this paper, we develop statistical and computational tools for the search for self-lensing flares in LSST. First, we use the python package \texttt{binlite} \cite{D’Orazio_2024} for toy-models of light-curves of eccentric, equal-mass SMBHBs. \texttt{binlite} incorporates accretion variability from hydrodynamical simulations of binaries and circumbinary disks, relativistic Doppler modulation, and self-lensing flares to generate synthetic light-curves. We also incorporate stochastic quasar DRW variability and downsample our light-curves at LSST-like cadences. In our initial attempt, we use standard LS periodograms, the most commonly used method for SMBHB searches, to recover the periodicity of the self-lensing flares. We show that even in ideal conditions (no noise, ideal sampling), the LS periodogram peak heights of periodic self-lensing flares cannot be robustly distinguished from that of DRW. This result is consistent with the fact that LS periodograms are known to struggle with detecting non-sinusoidal (sawtooth shape) periodicity \cite{lin2025}. Alternatively, we use matched filters to identify periodicities and to recover binary parameters from self-lensing flares even in the presence of DRW noise and LSST-like observation gaps. In order to avoid searching a wide parameter space, we develop a three-step procedure to zoom-in and sample the vicinity of the true injected binary parameters.

The rest of this paper is organized as follows. In \S~\ref{sec:Methods}, we describe our methodology, including a summary of \texttt{binlite}, simulations of DRW variability, and simulations of LSST noise and sampling.  In \S~\ref{sec:Results}, we present our attempt to use LS periodograms as well as the parameter recovery capability using matched filters. 
In \S~\ref{sec:discussion}, we discuss the current limitations and future directions for matched filter approaches using hydrodynamical simulations. In \S~\ref{sec:conclusion}, we summarize our main conclusions.

\section{Methods}
\label{sec:Methods}

In this section we outline the methods of generating realistic mock light-curves. We start from our models for binary AGN periodicity and then describe our implementation of DRW noise. Finally, we describe our adopted LSST instrumental noise and  irregular sampling. Both DRW and LSST white noise are assumed to be additive, so our final light-curve is \begin{equation}
    d(t) = s_{\rm binlite}(t)+n_{\rm DRW}(t)+n_{\rm LSST}(t),
\end{equation} where the \texttt{binlite} model is denoted $s_{\rm binlite}$ to emphasize that it is the signal we want to extract from the DRW and LSST noise.

\subsection{Model binary light-curves}\label{subsec: binlite}

\begin{figure*}
    \centering
    \includegraphics[width=\textwidth]{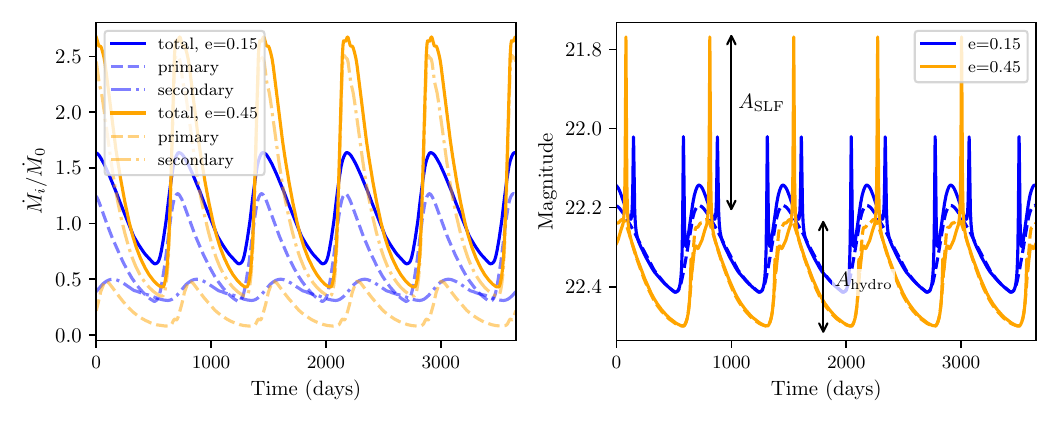}
    \caption{{\em Left panel}: Normalized accretion rates that were used to generate the two light-curves 
    in the right panel. The dashed and dash-dotted curves indicate the accretion from the 
    primary and the secondary, respectively. Note that in this example, the ratio of the time-averaged accretion rates $\langle \dot{M}_2(t)\rangle/\langle\dot{M}_1(t)\rangle$ is different for $e=0.15$ vs $e=0.45$, meaning that despite $M_1=M_2$ for \texttt{binlite}, the relative brightness of the two disks $F_{\nu,1}/F_{\nu,2}$ can be different for different eccentricities (See \S~\ref{subsec: binlite} for discussion).
    {\em Right panel}: Examples of \texttt{binlite}-simulated optical ($r$-band) light-curves 
    from equal-mass binaries with total mass $M = 10^9\,{\rm M}_\odot$, orbital period 
    $T_{\rm orb} = 2~\mathrm{yr}$ at redshift $z = 1$, accreting at $30\%$ of its 
    Eddington limit and inclined at $I = 88^\circ$. 
    We visualize the effects of varying the orbital eccentricity for two values: 
    $e = 0.15$ and $e = 0.45$. 
    The dashed curves indicate the light-curves assuming no self-lensing magnification 
    or Doppler boost effects (i.e. $\mathcal{M}_1 = \mathcal{M}_2 = 1$ in 
    Eq.~\ref{eq: total_flux}). 
    We also indicate the amplitude of two periodic binary signatures: self-lensing 
    ($A_{\rm SLF}$) and accretion variability ($A_{\rm hydro}$) for reference. }
    \label{fig:binlite-examples}
\end{figure*}

To simulate self-lensing flares and accretion variability, we use the python package \texttt{binlite} from \cite{D’Orazio_2024} (hereafter B24). We provide a brief overview of their methods below, but refer the reader to B24 for more details.

B24 consider a binary characterized by total mass $M=M_1+M_2$ where the masses of the two components are equal: $q=M_2/M_1=1$. The binary orbit has eccentricity $e\in (0,0.8)$, semi-major axis $a$, and orbital period $T_{\rm orb}$. A locally isothermal circumbinary disk (CBD) accretes onto the binary at a user-specified fixed rate of $\dot{M}_{\rm CBD}=\langle\dot{M}_1(t)+\dot{M}_2(t)\rangle$, and $Q\equiv\langle \dot{M}_2(t)\rangle/\langle\dot{M}_1(t)\rangle$ 
is defined as the ratio of time-averaged accretion rates between the accretion onto each component calculated from a hydrodynamical simulation. Note that the averaging is performed over dozens of orbits.

An important feature of \texttt{binlite} is that despite the two black holes having the same mass, their accretion rates, when averaged over dozens of orbits, differ ($Q\neq 1$).   This accounts for the behavior observed in hydrodynamical simulations: the two black holes are surrounded by a non-axisymmetric circumbinary disk, which typically precesses on time-scales corresponding to hundreds of orbits.  As a result, one black hole out-accretes the other, with their roles switching back-and-forth on this precession timescale.   A typical observation time (i.e. several years to a decade) is much shorter than this timescale, except for rare, ultra-compact binaries with periods of at most several days.   In practice, this implies that during an LSST observation, an equal-mass binary can be observed with a range of possible long-term average accretion-rate ratios $Q$.  Furthermore, the probability distribution of this average $Q$ depends on eccentricity.   In our work, we include the dependence of $Q$ on $e$, as provided by \texttt{binlite}. This assumes, for simplicity, that $Q=Q(e)$ is deterministic, i.e. yields a single $Q$, roughly corresponding to the value at which the binary spends most of its time.   Future analyses should incorporate the full probabilistic nature of $Q$ and allow
for a range of possible $Q$ values for each eccentricity.

The above fixes the mean accretion rate ratio.   On top of this, the variation of the accretion rates on short (orbital) timescales are included as
\begin{equation}
    p(t)=\frac{\dot{M}_1(t)}{\langle\dot{M}_1(t)+\dot{M}_2(t)\rangle}, \;\;s(t)=\frac{\dot{M}_2(t)}{\langle \dot{M}_1(t)+\dot{M}_2(t)\rangle}.
\end{equation} 
Here, $\dot{M}_1(t)$ and $\dot{M}_2(t)$ are Fourier reconstructions of accretion rates in 2D hydrodynamical simulations and thus are deterministic and periodic. This reconstruction is valid as a close approximation to the original accretion rates as the power of the Fourier components is concentrated at integer multiples of the orbital frequency, as shown in the periodograms of B24's Fig 1. A comparison, with close match, between the simulation and the Fourier reconstructions can be found in B24's Fig 2. 

Next, each BH component of the binary is assumed to have a minidisk characterized by the thin-disk temperature profile $T(r)\propto (M_i\dot{M}_{i, \rm avg} / r_i^3)^{1/4},$ where $M_i, \dot{M}_{i, \rm avg}=\langle \dot{M}_i\rangle$ and $r_i$ are the masses, time-averaged accretion rates and the distance to each BH, respectively. The CBD has the same temperature distribution but is assumed to be centered around a BH with total mass of $M=M_1+M_2$ located at the center of mass of the system. Each minidisk is assumed to extend from its innermost stable circular orbit (ISCO) $r_{\rm ISCO, i}(M_i)=6GM_i/c^2$ to the tidal truncation radius $r_{\rm tidal, i}=0.27a,$ where $a=(GMT^2/4\pi^2)^{1/3}$. The CBD is assumed to extend from $2a$ to $100a.$ Finally, assuming the two minidisks and the CBD are optically thick and their spectra can be approximated as blackbodies, the average fluxes from each disk are identically calculated as: \begin{equation}\label{eq:disk-flux}
    F_{\nu,i}=\frac{2\pi \cos I}{d^2}\int B_\nu[T_i(r_i)]r_idr_i,
\end{equation} for a binary located at luminosity distance $d=d(z)$, where $z$ is the cosmological redshift, $B_\nu$ is the Planck function and $I$ is the binary inclination where we use the convention that $90^\circ$ is edge-on and $0^\circ$ is face-on. The factor of $\cos I$ indicates that we assume isotropic emission and that only the perpendicular component of the flux from the disk will be observable. The binary orbital plane and the two minidisks are assumed to be co-planar, given that the alignment of the accretion planes of SMBHs with the orbital axis is expected to be efficient \cite{Bogdanovic_2007}. 

Finally, the total observed flux from the system is\begin{equation}\label{eq: total_flux}
    F_\nu = F_{\nu, 1}\mathcal{M}_1(t) p(t)+F_{\nu, 2}\mathcal{M}_2(t) s(t) + F_{\nu, \rm CBD},
\end{equation} where $F_{\nu,1}, F_{\nu,2}, F_{\nu,{\rm CBD}}$ are the average fluxes from the primary, secondary, and CBD respectively and $\mathcal{M}_i(t)$ is the total modulation from self-lensing magnification and Doppler boost, and we have explicitly noted the time dependence. The self-lensing magnification of the flux is calculated assuming the point source approximation \begin{equation}
    \mathcal{M}(u)=\frac{u^2+2}{u\sqrt{u^2+4}},
\end{equation} where $u$ is the angular separation of the binary in units of the Einstein radius. All of the features introduced above in this section, with the final product being a total observed flux time series, can be obtained by generating a \texttt{BinaryAlphaDisk} object and using the \texttt{periodic\_flux\_series\_from\_bad} function. After obtaining the flux series (given in cgs units) from \texttt{binlite} we convert to AB magnitudes. In section \ref{subsec: LSST-sims} we take this flux series $F_\nu$ and simulate it with LSST-like photometric noise and observation gaps before converting to magnitudes. The statistical analysis using LS periodograms and Matched Filters was done in magnitudes.

\subsection{Intrinsic quasar variability}

Meanwhile, we model the intrinsic optical variability of quasars as a damped random walk (DRW) \cite{MacLeod_2010}, which has a typical characteristic timescale $\tau_{\rm DRW}$ on the order of a few hundred days and can mimic periodic signals for a typical survey length, giving the appearance of a binary \cite{Vaughan+2016}. The DRW is characterized by an exponential covariance matrix: \begin{equation}\label{eq: drw covariance matrix}
    S_{ij}= A_{\rm DRW}^2\exp[-|t_i-t_j|/\tau_{\rm DRW}].
\end{equation} Here $A_{\rm DRW}$ (typically denoted as $\sigma_{\rm DRW}$ in other works, but denoted differently in this work for direct comparison to $A_{\rm SLF}$) is the long-term ($\Delta t\gg \tau_{\rm DRW}$) DRW variance in magnitude. $\tau_{\rm DRW}$ is a characteristic observer-frame damping timescale and $t_i,t_j$ are two observation times of the light-curve. We used the Python package \texttt{astroML}'s \texttt{generate\_damped\_RW} function to generate mock DRW light-curves. 

\subsection{LSST simulation}\label{subsec: LSST-sims}
The simulated SMBH binary light-curves are then downsampled on the cadence of the Rubin Observatory LSST \texttt{baseline\_v3.3\_10yrs} OpSim run~\cite{rubin_baseline_v3_3}. LSST-like instrumental noise is calculated using the semi-analytic model outlined in the publicly released \texttt{LSE-40, LSST Photon Rates and SNR Calculations, v1.2} document~\cite{ivezic2010lsstsnr}, of which we give a brief overview. 

Given a source flux $F_\nu(\lambda)$, the flux transmitted through the atmosphere to the telescope pupil is given by \begin{equation}
    F_{\nu}^{\rm pupil}(\lambda)=F_\nu(\lambda)S^{\rm atm}(\lambda),
\end{equation} where $S^{\rm atm}$ is the probability that a photon with wavelength $\lambda$ will reach the telescope. Then given $F_{\nu}^{\rm pupil},$ the number $C_b$ of recorded analog-to-digital units (ADUs) in a given band $b$ is given by \begin{equation}\begin{split}\label{eq: flux to ADU}
    C_b=C\int_0^\infty F_\nu^{\rm pupil}(\lambda) S_b^{\rm sys}(\lambda)/\lambda d\lambda\\ \approx CF_\nu(\lambda_{\rm eff, b})\int_0^\infty S^{\rm atm}(\lambda)S^{\rm sys}_b(\lambda)/\lambda d\lambda,
\end{split}
\end{equation} where the effective wavelength in each band $b$ is calculated as $\lambda_{\rm eff, b}=\int\lambda S_b^{\rm sys}(\lambda)d\lambda / \int S_b^{\rm sys}(\lambda)d\lambda$, $S^{\rm sys}_b(\lambda)$ is the probability that a photon at the telescope pupil will be converted into a ADU count, and the conversion from flux to ADU count is given by $C=A\Delta t_{\rm exp}/gh$, where $A=\pi(321.15)^2 {\rm cm}^2$ is the mirror area, $\Delta t_{\rm exp}=30 s$ is the average exposure time, $h=6.626\times 10^{-27} {\rm erg\times s}$ is Planck's constant, and $g=2.2$ is the number of photo-electrons per ADU count.
We incorporate the sky background flux density $B_\nu [{\rm Jy/arcsec^2}]$ provided in the OpSim data by converting the flux density to a ADU count per pixel \begin{equation}
    B_b=CB_\nu(\lambda_{\rm eff, b})\int_0^\infty S^{\rm atm}(\lambda)S^{\rm sys}_b(\lambda)/\lambda d\lambda \cdot \left(\frac{0.2 {\rm arcsec}}{\rm pixelScale}\right)^2.
\end{equation} In LSST, point-like sources will be optimally collected over $n_{\rm \rm eff}=2.266 ({\rm FWHM}/{\rm pixelScale})^2$ pixels, where we adopt the simulated data in OpSim for atmospheric seeing (FWHM). 

Finally, we take the instrumental noise per pixel (dark current, read noise from sensors and electronics) to be $\sigma_{\rm instr}^2=12.7$ electrons per pixel.

Combining all sources of noise, we can calculate the signal-to-noise ratio (SNR) of each data point in our simulation as \begin{equation}\label{eq: LSST SNR}
    {\rm SNR}=\frac{C_b}{\sqrt{\frac{C_b}{g}+(\frac{B_b}{g}+\sigma_{\rm instr}^2)n_{\rm eff}}}.
\end{equation} We can then calculate the LSST noise in magnitude units as \begin{equation}
    n_{\rm LSST}\; ({\rm mag}) =\left|\frac{dm}{dF_\nu}\right|\Delta F_\nu=\frac{2.5}{{\rm ln}10}\frac{\Delta F_\nu}{F_\nu}\sim \frac{1.09}{\rm SNR}.
\end{equation}

To summarize, we incorporate LSST-like noise $n_{\rm LSST}(t)$ and LSST-like cadences from the \texttt{baseline\_v3.3\_10yrs} OpSim run in our mock light-curves.

\section{Results}
\label{sec:Results}

In order to detect the periodicity in the mock light-curves, we initially attempt using the most common technique in the literature, i.e. LS periodograms. We inject self-lensing flares into a DRW and see if a LS periodogram will produce a peak that is located near the true binary periodicity and if the peak height is high enough that it cannot be mimicked by pure DRWs.

\subsection{LS periodograms: SLF+DRW vs DRW}\label{subsec: slf vs drw}
\begin{figure*}
    \centering
    \includegraphics[width=1\linewidth]{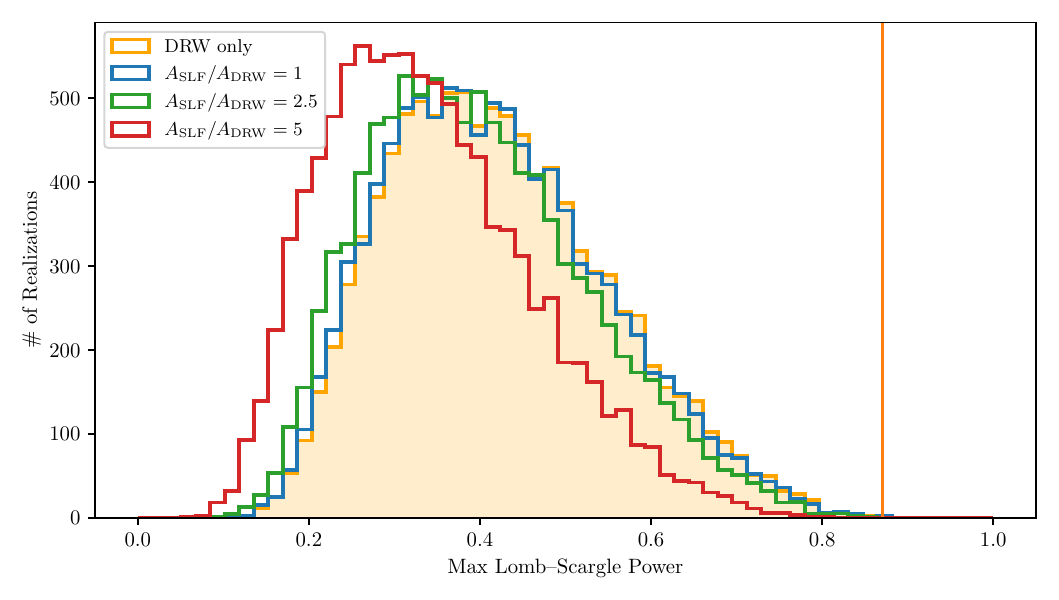}
    \caption{Histograms of the maximum peak powers in 10,000 DRW (orange) realizations and in corresponding SLF+DRW (blue, green, red) light-curves. The height of the self-lensing flare $A_{\rm SLF}$ was varied manually so that it was $1, 2.5, 5$ (blue, green, red respectively) times larger than the DRW amplitude $A_{\rm DRW}.$ The maximum peak height value among the 10,000 pure DRW cases is marked by a vertical orange line. Out of the SLF-included light-curves, none have peak heights greater than this maximum. Counterintuitvely, as $A_{\rm SLF}/A_{\rm DRW}$ is increased, the LS power decreases (see Fig.~\ref{fig:individual light curves} for a visual explanation of this effect).}
    \label{fig:SLF_vs_drw_main}
\end{figure*}

We begin by introducing a fiducial binary, whose parameters are summarized in Table~\ref{table:fiducial_binary}. We chose a relatively long orbital period and large total mass, so that the lensing duration will be long but multiple orbits will be present in the light-curve. We also chose an eccentricity which is expected from hydrodynamical simulations \cite{Siwek_2023}. We restrict our analysis to equal-mass binaries, consistent with the simulations of B24. We additionally select a relatively nearby and bright system (in terms of redshift and Eddington ratio) and adopt typical DRW parameters \cite{MacLeod_2010}. We simulate the photometric variability of this binary over 10 years to mimic a light-curve expected in LSST.

\begin{table}
\centering
\begin{tabular}{@{}lc@{}}
\toprule
\multicolumn{2}{c}{\textbf{Fiducial Binary}} \\
\midrule
\textbf{Binary Parameter} & \textbf{Injected} \\
\midrule
$T_{\rm orb}$ (years) & 1 \\
$e$ & 0.4 \\
$I$ & $88^\circ$ \\
$\omega$ & $90^\circ$ \\
$q=M_2/M_1$ & 1 \\
$\log(M/{\rm M}_\odot)$ & 9 \\
$z$ & 0.5 \\
$f_{\rm edd}$ & 0.2 \\
\midrule
\multicolumn{2}{l}{\textbf{DRW Parameters}} \\
\midrule
$A_{\rm DRW}$ (mag) & 0.1 \\
$\tau_{\rm DRW}$ (days) & 300 \\
\midrule
\multicolumn{2}{l}{\textbf{light-curve Settings}} \\
\midrule
Total Duration & 10 years \\
Filter & LSST r-band \\
\bottomrule
\end{tabular}
\caption{
Binary parameters, DRW parameters, and light-curve settings of the fiducial binary. The resulting self-lensing flares have height of $A_{\rm SLF}\sim 0.5$ mag.}
\label{table:fiducial_binary}
\end{table}

\begin{figure*}
    \centering
    \includegraphics[width=1\linewidth]{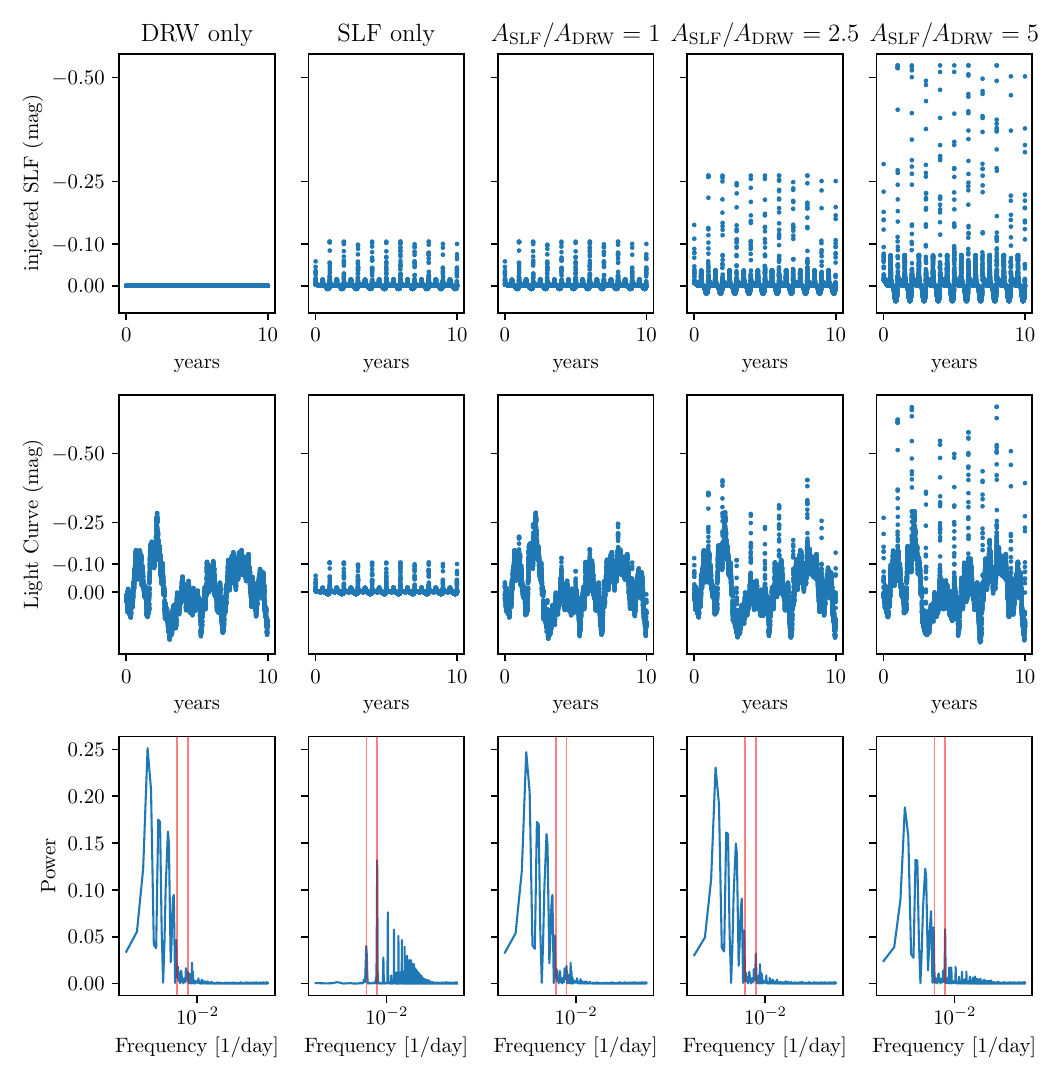}
    \caption{light-curves and their LS periodograms representative of the trends in Fig.~\ref{fig:SLF_vs_drw_main}. Each column varies $A_{\rm SLF}/A_{\rm DRW}$, where the first column shows pure noise, the second column shows pure SLF signal, and the final three columns show scenarios where the signal to noise ratio $A_{\rm SLF}/A_{\rm DRW}$ is increased from 1, 2.5, 5. The same DRW noise realization is kept throughout the plots. The top row shows the injected signal of each light-curve and the light-curves are shown in the middle row. The bottom row shows LS periodograms of the light-curves, where the two vertical red lines show the injected orbital frequency $f_{\rm orb}=1/T_{\rm orb}$ and $2f_{\rm orb}$ in units of 1/day. As $A_{\rm SLF}/A_{\rm DRW}$ increases, the DRW peaks decrease and the SLF peaks increase, but the SLF peaks never become the highest peaks.}
    \label{fig:individual light curves}
\end{figure*}

To analyze the periodicity of self-lensing flares relative to DRW, we follow a simplified version of the analysis used in \cite{Charisi+2016}, using a LS periodogram to compare periodogram peak heights between an injected signal with a DRW noise added relative to a pure DRW noise light-curve. We only consider the periodic variability in magnitude units caused by self-lensing, given that LS periodograms are known to struggle with non-sinusoidal signals \cite{lin2025}, and manually add in a damped random walk. We neglect any hydrodynamical variability and noise/gaps from LSST. The injected binary, DRW parameters, and the light-curve settings are shown in Table~\ref{table:fiducial_binary}.

We identically downsample a 10-year SLF+DRW light-curve and DRW-only light-curve at 10,000 random times, using the same DRW realization, and compute the LS periodograms for each scenario. We repeat this process for the same SLF, but with 10,000 DRW realizations. For each pair (SLF+DRW vs DRW) of LS periodograms, we recorded the frequency and height of the highest peak. We present these distributions in Fig.~\ref{fig:SLF_vs_drw_main}, where we also show the results for different $A_{\rm SLF}/A_{\rm DRW}$ ratios, varied by dividing through the fiducial $A_{\rm SLF}\sim 0.5$ mag manually with a constant factor of 2.0 or 5.0, and keeping $A_{\rm DRW}=0.1$ mag constant. 

Finally, we checked one criterion from the periodograms, whether the periodogram power of the highest peak was greater than 10,000 DRW realizations, shown in a vertical orange line in Fig.~\ref{fig:SLF_vs_drw_main}. This criterion is generous as it only guarantees false alarm probabilities of $10^{-4}$ while there are millions of quasars expected in LSST. However, we find that even this criterion is not satisfied, i.e. this analysis is not a reliable method of detecting flares even for this idealized light-curve. In all scenarios the number of SLF-included light-curves satisfying this criterion is 0 out of 10,000, which can also be seen in Fig.~\ref{fig:SLF_vs_drw_main} by comparing the orange (DRW-only) distribution vs the red, green, blue SLF-included distributions. In Fig.~\ref{fig:SLF_vs_drw_main}, increasing the $A_{\rm SLF}/A_{\rm DRW}$ ratio seems to counterintuitively shift the peak-height distributions to lower heights.

To understand these trends, we show a typical light-curve realization in Fig.~\ref{fig:individual light curves}. In the first (leftmost) periodogram, there is an aliasing peak of height $\sim 0.25$ caused by the DRW realization, which is higher than the injected SLF's tallest peak height of $\sim 0.13$ in the second periodogram. Note that this DRW periodogram peak height is typical, as can be seen from the orange DRW distribution from Fig.~\ref{fig:SLF_vs_drw_main}. This aliasing peak at $f_{\rm false}\sim 4\times 10^{-4}$ [${\rm days}^{-1}$] renders the true $f_{\rm orb}, 2f_{\rm orb}$ peaks undetectable. For larger $A_{\rm SLF} / A_{\rm DRW},$ the false peaks from DRW progressively decrease, and the true $f_{\rm orb}, 2f_{\rm orb}$ peaks increase, but do not emerge as the highest peak even for $A_{\rm SLF} / A_{\rm DRW}=5$ (as shown in the rightmost periodogram).

\begin{figure}
    \centering
    \includegraphics[width=1\linewidth]{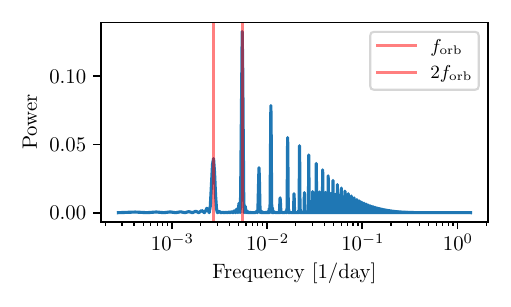}
    \caption{Fourier transform of the injected self-lensing flares in the second column of Fig.~\ref{fig:individual light curves}. There was no noise included and the flares were sampled at a regular and dense cadence (opposed to random cadence in Fig.~\ref{fig:individual light curves}) for this illustration.}
    \label{fig:slf fourier transform}
\end{figure}

To isolate sampling effects, we also show the Fourier transform of the SLF-only light-curve (second column of Fig~\ref{fig:individual light curves}) in Fig.~\ref{fig:slf fourier transform}. The Fourier transform is the LS periodogram given regularly spaced and well-sampled data and even in this case the maximum peak height is only $\sim 0.13.$ Thus we conclude that a naive application of the LS periodogram fails to detect SLFs \cite{lin2025}. 
This conclusion is very similar to the recent work by \cite{lin2025}, who found that sawtooth-shaped hydrodynamical periodicity is difficult for LS periodograms to pick up.  In both cases, this failure is caused by the non-sinusoidal pulse shape, which results in the power moving away from the single orbital frequency, and spread over many Fourier modes.
In principle, these non-sinusoidal pulse shapes
could be modeled an incorporated into a LS analysis, but we instead turn to matched filters as an alternative approach.

\subsection{Matched filters}

The problem of recovering SLFs involves identifying not just a single prominent Fourier peak, but a collection of Fourier peaks at the orbital frequency and its harmonics (Fig.~\ref{fig:slf fourier transform}). We propose matched filters as an alternative to LS periodograms for this problem, with the additional advantage that we can infer binary parameters, unlike with LS periodograms or in many other period-detection algorithms.

\subsubsection{Matched filters: approach}

\begin{figure}
    \centering
    \includegraphics[width=\linewidth]{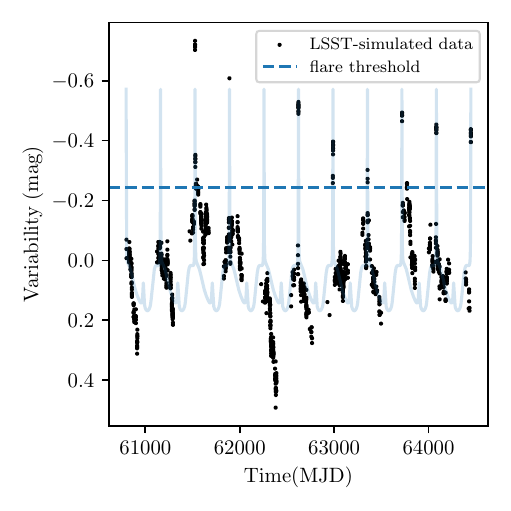}
    \caption{Illustration of the mock light-curve of the fiducial binary with realistic LSST data quality. The faint blue light-curve represents the \texttt{binlite} binary-template used to generate the signal $s(\hat{\theta})$ and the black data points $d(t)=s(\hat{\theta})+n(t)$ represent the downsampled binary-template at LSST-like observation times with LSST-like noise and a DRW added. An estimate for the flare threshold is drawn with a dashed blue line,  calculated via a simple analytical approximation for the lensing duration in Eq.\ref{eq:lensing duration}.}
    \label{fig:fiducial light curve}
\end{figure}

With matched filters, we aim to analyze the light-curve of the fiducial binary, shown in Fig.~\ref{fig:fiducial light curve} where the binary-template has been degraded by adding DRW and LSST noise, and sampled at LSST-like observation times. The realistic mock light-curve of the fiducial binary is shown with black points in Fig.~\ref{fig:fiducial light curve}, and the faint blue light-curve shows the \texttt{binlite} light-curve used to generate the mock light-curve. We also assume that from other measurements, we have found the redshift $z$ and total mass $M=M_1+M_2$ of the quasar to infinite precision. In addition, having a 10-year light-curve at the end of LSST will determine a mean luminosity, which then determines an Eddington ratio for a given mass $M$-- which again we will assume we know precisely. Under these assumptions, the binary is characterized by the four free parameters of orbital period $T_{\rm orb}$, eccentricity $e$, binary inclination $I$, and the argument of periapsis $\omega$ (which is defined so that $\omega=0^\circ$ means the line of sight is parallel to the semi-minor axis and $\omega=90^\circ$ means the line of sight is parallel to the semi-major axis). We assume that these parameters are unknown, and have priors summarized in Table~\ref{table:fiducial_binary_priors}. 

\begin{table}[h]
\centering
\begin{tabular}{@{}lcc@{}}
\toprule
\multicolumn{3}{c}{\textbf{Fiducial Binary}} \\
\midrule
\textbf{Binary Parameter} & \textbf{Injected} & \textbf{Prior} \\
\midrule
$T_{\rm orb}$ (years) & 1 & $U[0.5,5]$ \\
$e$ & 0.4 & $[0.1, 0.2, 0.3, \ldots, 0.7, 0.79]$ \\
$I$ & $88^\circ$ & $\cos^{-1}(U[0,1])$ \\
$\omega$ & $90^\circ$ & $U[0^\circ,360^\circ]$ \\
\bottomrule
\end{tabular}
\caption{
The priors of the four free parameters in this matched filter exercise and the true injected values that we attempt to recover via matched filters. Any parameters not mentioned in this table ($q,M, z, f_{\rm edd}$, $A_{\rm DRW}$, $\tau_{\rm DRW}$) are assumed to be known and are taken directly from Table~\ref{table:fiducial_binary}. $U[a,b]$ denotes a uniform distribution in the range $[a,b].$}
\label{table:fiducial_binary_priors}
\end{table}

\begin{figure*}
    \centering
    \includegraphics[width=\linewidth]{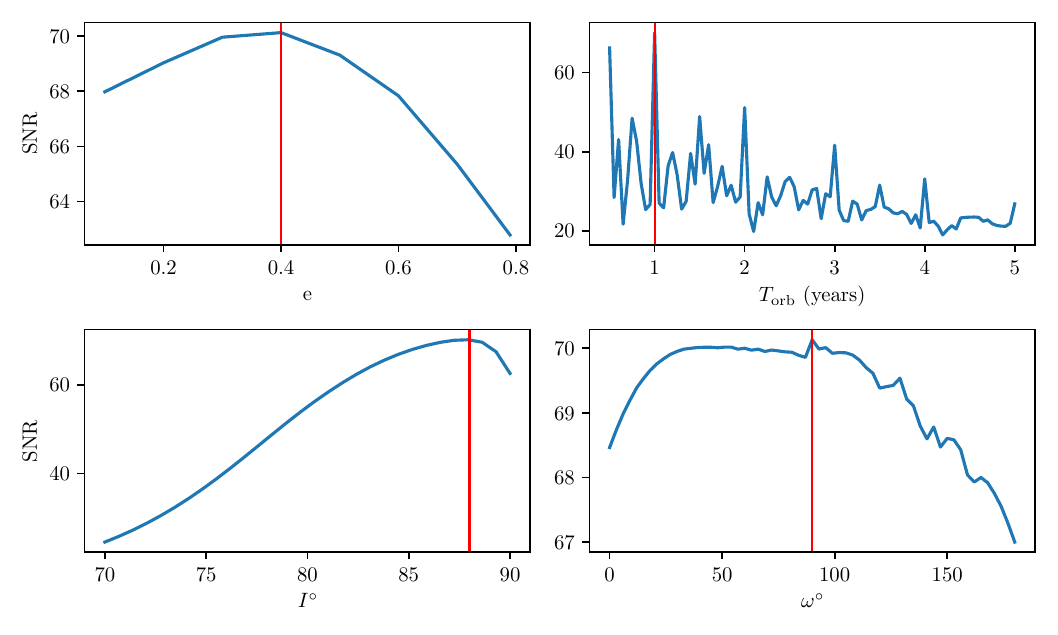}
    \caption{Matched filter SNR, maximized over a phase shift $\Delta$, of the fiducial binary relative to binary-templates in which 3 out of 4 binary parameters are fixed. The true binary parameters $e=0.4, I=88^\circ, T_{\rm orb}=1{\rm yr}, w=90^\circ$ are marked with red vertical lines. {\em Top left panel}: eccentricity dependence, i.e. $\rho(e, T_{\rm orb}=1{\rm yr}, I=88^\circ, w=90^\circ)$. {\em Top right panel}: orbital period dependence--there are significant false peaks at $0.5T_{\rm orb}, 2.0T_{\rm orb}, 3.0T_{\rm orb}$ etc. {\em Bottom left panel}: binary inclination dependence. {\em Bottom right panel}: argument of periapsis dependence-- there is only a mild dependence on $\omega,$ with SNRs of 67 and above being achieved as long as the other orbital parameters are fixed. The true injected parameters have the highest overall SNR of 71.}
    \label{fig:SNR dependences}
\end{figure*}

The matched filter is a method used to detect weak signals embedded in noise by maximizing the signal-to-noise ratio \begin{equation}\label{eq: MF SNR}
    \rho(\theta)\equiv \frac{s(\theta)*d}{\sqrt{s(\theta)*s(\theta)}},
\end{equation} which is the normalized inner product between a binary-template time series $s(\theta)$ (i.e. $s_{\rm binlite}(\theta)$) where $\theta$ are the template binary parameters, $\hat{\theta}$ are the true binary parameters that create the binary signal and observed time series $d(t)=s(\hat{\theta})+n(t)$, where $n(t)=n_{\rm DRW}(t)+n_{\rm LSST}(t)$ with covariance $C_{ij}=S_{ij}+N_{\rm LSST},$ where $C_{ij}$ is a square matrix whose length is the number of light-curve data points, $S_{ij}$ is the DRW covariance from Eq.\ref{eq: drw covariance matrix} and $N_{\rm LSST}$ is a diagonal matrix with (uncorrelated) LSST noise. The binary-template $s(\theta)$ is interpolated and sampled at the observation times of the mock light-curve $d(t)$. We also maximize the SNR up to a time shift, so the numerator of Eq.\ref{eq: MF SNR} is the maximum SNR out of all possible shifts $\Delta$ of the binary-template light-curve: 
\begin{equation}
    s(\theta) * d
    = \max\limits_{\Delta} \left( \sum_t s[\Delta + t]^T C^{-1} d[t] \right).
\end{equation}
 The denominator is calculated as \begin{equation}
    \sqrt{s(\theta)*s(\theta)}=\sqrt{s(\theta)^T C^{-1}s(\theta)},
\end{equation} which prevents bias towards binary-templates with poor fits but large amplitudes. In addition, we subtract the mean from the mock data $d(t)$ and the templates $s(\theta)$ before these calculations for the same purpose of preventing biases towards templates with large amplitudes. In practice, for a given template with known orbital period $T_{\rm orb}'$ we shift a given light-curve from 0 to $T_{\rm orb}'$ in 1.0 day increments to find the maximum SNR.

We note that due to large LSST observation gaps, the Fourier transform of the data $\tilde{d}$ cannot be computed without some degree of error, especially without some interpolation scheme. Thus we opt to compute the cross-correlation Eq.\ref{eq: MF SNR} in the time domain and not the frequency domain-- although this method would be computationally more efficient through the use of Fast Fourier Transforms. 

To efficiently match a large number of binary-templates to the mock light-curve, we pre-compute \texttt{binlite} binary-templates on the parameter space (the "template bank" method in gravitational wave analysis) and for a given light-curve, we calculate the SNR $\rho$ for all the templates and look at the templates which have the highest SNR $\rho$. This has the advantage that these binary-templates are reusable for any number of light-curves that one wishes to analyze, as opposed to a method like Monte Carlo Markov Chain (MCMC), which requires template generation or interpolation during a search. However, this introduces a dependence on the spacing of the grid of binary-templates in parameter space, where a sparse grid will lead to missing the SNR-maximizing template, but an overly dense grid will become exponentially difficult to store and process.

Next, from the priors described in Table~\ref{table:fiducial_binary_priors}, we choose 10000 points according to a Sobol sequence \cite{sobol1967} using {\tt Scipy}'s {\tt Sobol} package and calculate the r-band \texttt{binlite} light-curve for each point. A DRW realization $n_{\rm DRW}(t)$ with DRW amplitude $A_{\rm DRW}$ of 0.1 mag and observed damping timescale $\tau_{\rm DRW}$ of 300 days was added to the binary signal and the signal+noise light-curve was sampled at LSST-like observation cadences and gaps, and LSST-like noise $n_{\rm LSST}(t)$ was also applied. For our fiducial binary at $\sim 20$ mag, the $1\sigma$ of LSST white noise is typically $0.01-0.05$ mag. In Fig.~\ref{fig:fiducial light curve}, we show the \texttt{binlite}-simulated light-curve in blue and the noise-degraded LSST light-curve with black points. The mean of the light-curve was subtracted off to clearly visualize the various variability (hydrodynamical, self-lensing, DRW).

\subsubsection{Matched filters: results}

We show 1-D slices of the distribution of SNR $\rho(\theta)$ in the parameter space in Fig.~\ref{fig:SNR dependences}, i.e. $\rho(e, T_{\rm orb}=1{\rm yr}, I=88^\circ, w=90^\circ)$ vs $e$, $\rho(I, T_{\rm orb}=1{\rm yr}, e=0.4, w=90^\circ)$ vs $I$, etc. We again mark the values of the injected binary parameters $\hat{\theta}$ with red vertical lines. For $T_{\rm orb}$, we find narrow SNR peaks at $0.5T_{\rm orb}$ and $2T_{\rm orb}$ and other SNR peaks $1.5T_{\rm orb}$ and $2.5T_{\rm orb}$. It is notable that these peaks are extremely narrow, of order few $\%$ of $T_{\rm orb}$. This can be attributed to the fact that the maximum SNR will be achieved when all the flares are "matched" in a given binary template, and a binary-template with an orbital period that is too far away from $T_{\rm orb}$ or one of its multiples will necessarily miss some of the 10 lensing flares of the fiducial binary. 

For the argument of periapsis $\omega$, if all the orbital parameters are fixed, changing $\omega$ will impact the timing of the flares. However, a shift of the template can account for this, leading to relatively small variations $\sim 3$ in the SNR, caused by small deviations in the fit to the mock light-curve, and the fact that we used shift spacings of $\Delta =1.0$ day.

For the orbital eccentricity, we find that there is only a mild loss in the SNR even for a template with a much different eccentricity. This is because for a template with the correct orbital period, inclination and argument of periapsis, the flares are already occurring at the correct times and (roughly) correct heights. We note that we use a discrete prior $e\in [0.1, 0.2, \cdots]$ because of how the accretion rates from the B24 hydrodynamical simulations are sampled as a function of eccentricity due to their sweep technique they employ. We discuss this issue in \S~\ref{sec:discussion}.

Finally, for the orbital inclination, which primarily determines the angular separation between the two BHs at closest alignment, shows a significant loss of SNR when $I\lesssim 80^\circ$, which will not produce flares. In the other direction, we also find that there is a loss of SNR when $I=90^\circ$, which should produce flares that are higher compared to $I=88^\circ.$ 

With an understanding of the expected SNR distribution in parameter space, and the SNR ($\rho_{\rm max}=71$) at the true parameters for the fiducial binary, we perform a three-step analysis procedure to recover the injected binary-template:
\begin{itemize}
    \item Step~\#1: Draw 10,000 binary-templates from the wide prior described in Table~\ref{table:fiducial_binary_priors}, and compute the matched-filter SNR for each. 
    \item Step~\#2: Delete the self-lensing flares from the data and repeat Step~\#1 to accurately determine the orbital period. Steps~\#1 and~\#2 together reduce the search space in $T_{\rm orb}$ and $I$. 
    \item Step~\#3: Draw 10,000 binary-templates from the narrowed prior in $T_{\rm orb}$ and $I$ and compute the SNRs.
\end{itemize}

\subsubsection{Step \#1: course grid}

\begin{figure}
    \centering
    \includegraphics[width=\linewidth]{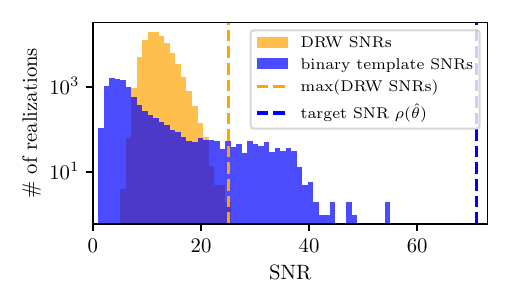}
    \caption{Distribution of the SNR $\rho$ when matching the mock light-curve $d(t)$ to $10^4$ sparsely sampled binary-templates and $10^5$ DRW realizations. The maximum SNR of the DRW realizations, 25, is noted with a dashed orange vertical line and the SNR at the injected values, 71 (see Fig.~\ref{fig:SNR dependences}), is noted with a dashed blue vertical line. We find that 536 templates out of the sparsely sampled $10^4$ binary-templates have SNRs greater than the maximum SNR from DRW templates.}
    \label{fig:null snrs}
\end{figure}

\begin{figure*}
    \centering
    \includegraphics[width=\linewidth]{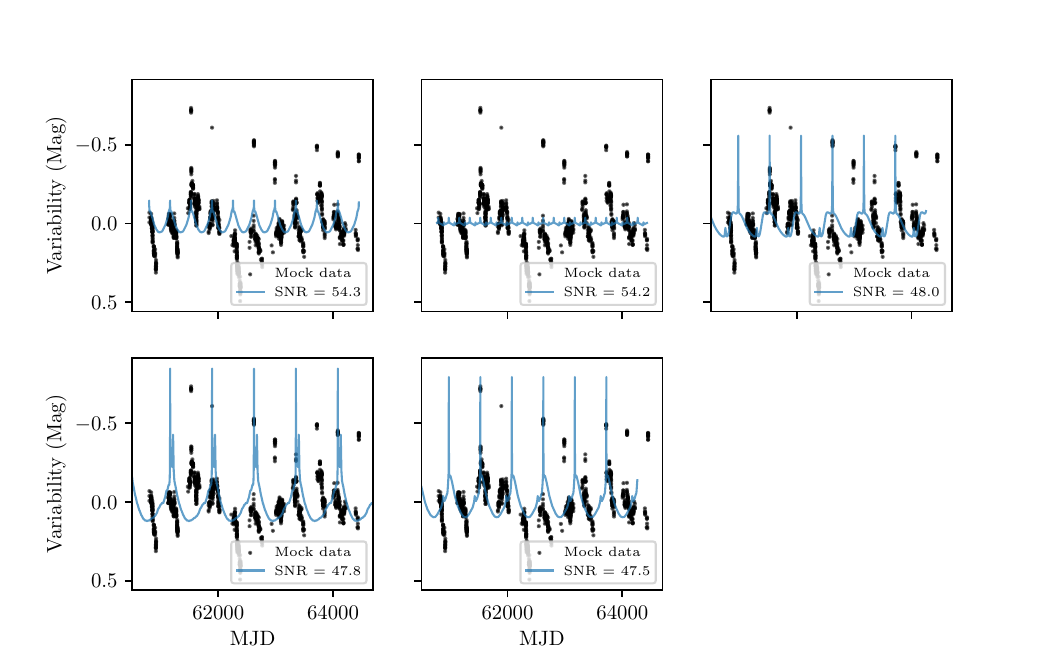}
    \caption{The five binary-templates with highest SNR from Step \#1, Fig.~\ref{fig:null snrs}, overlaid on the mock fiducial light-curve. In order of higher SNR to lower SNR, the binary-templates have orbital periods of $T_{\rm orb}=1.00, 0.50, 1.50, 2.00, 1.50$ years. }
    \label{fig:optimal_shift}
\end{figure*}

\begin{figure*}
    \centering
    \includegraphics[width=\linewidth]{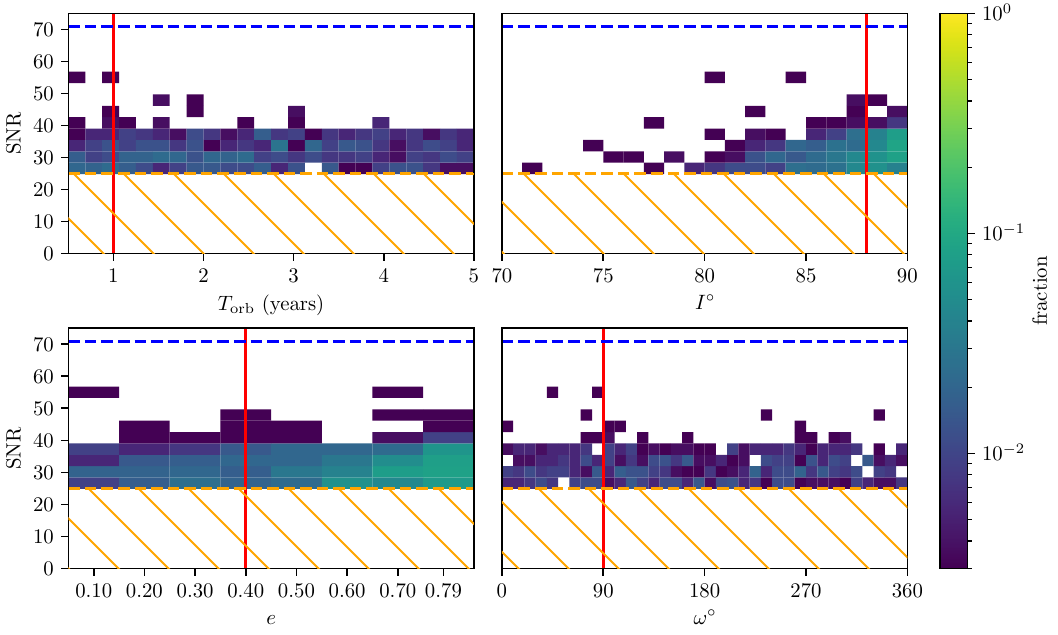}
    \caption{Step \#1, 2D histogram of 536 out of 10,000 binary-templates with parameters sampled from a wide prior, binned by their binary parameters (horizontal axis) and matched filter SNRs (vertical axis). The color of each (parameter, SNR) bin indicates the fraction of the selected 536 binary templates that fall into that bin. These binary-templates have SNRs higher than the maximum SNR achieved with $10^5$ DRW realizations from Fig.~\ref{fig:null snrs}, marked with an orange dashed line. While our goal was to sample the SNR peak at the injected parameters ($\sim 71$, marked with blue dashed line), we find that most binary-templates have SNRs of 25-40 when matched to the mock data. Injected binary parameters are marked with red lines.}
    \label{fig:step1_histograms}
\end{figure*}

We present the results of the SNRs of the 10,000 binary-templates in Fig.~\ref{fig:null snrs}. The SNRs should be compared to the null hypothesis of assuming there is no binary in the light-curve. Thus we also generate $10^5$ pure DRW noise realizations with identical parameters ($A_{\rm DRW}=0.1, \tau_{\rm DRW}=300$d) to the ones used to generate the mock light-curve and compute their matched filter SNRs relative to the mock data. We find that DRWs can mimic binary signals up to an SNR of $25.$ This maximum value is shown with an orange dashed vertical line. We also find that 536 out of 10,000 binary-templates have SNRs that cannot be reached by these noise realizations. We show 5 binary-templates (after optimizing their time shift) with the highest SNR out of these 536 templates in Fig.~\ref{fig:optimal_shift}. Although the templates match to several of the flares, none of them match to all of the flares present in the data. This can also be seen numerically, where the best-fit templates have SNRs of 47-54 while a best-fit close to the correct model should have a SNR close to 71 (see Fig.~\ref{fig:SNR dependences}). The five binary-templates shown in Fig.~\ref{fig:optimal_shift} have binary parameters and optimal template shifts summarized in Table~\ref{table:step1_bestfits}. We find that for these five examples, the orbital periods are within 1\% of either 1, 1.5, 2 years and all binary-templates had inclinations of at least 80 degrees.

\begin{table}[h]
\centering
\begin{tabular}{@{}ccccccc@{}}
\toprule
Rank & SNR & $T_{\rm orb}$ (yr) & $I(^\circ)$ & $e$ & $\omega(^\circ)$ & $\Delta_{\rm opt}$ (d) \\
\midrule
Injected & 70.6 & 1.00 & 88.00 & 0.40 & 90.00  & 0   \\
\midrule
1 & 54.3 & 1.00 & 80.96 & 0.70 & 84.54  & 2   \\
2 & 54.2 & 0.50 & 84.15 & 0.10 & 40.07  & 23  \\
3 & 48.0 & 1.50 & 88.01 & 0.40 & 70.25  & 376 \\
4 & 47.8 & 2.00 & 87.27 & 0.79 & 333.67 & 353 \\
5 & 47.5 & 1.50 & 87.56 & 0.70 & 230.37 & 376 \\
\bottomrule
\end{tabular}
\caption{
Top-ranked binary-templates from Step~\#1 (Fig.~\ref{fig:null snrs}). 
First row shows the injected parameters for the fiducial binary and each row below lists the ranking (by SNR), SNR value, and corresponding binary parameters, including the optimal template shift $\Delta_{\rm opt}$ which maximizes the SNR.
}
\label{table:step1_bestfits}
\end{table}

We also show the parameter and SNR distribution of these binary-templates in a 2D histogram in Fig.~\ref{fig:step1_histograms}, again noting that while many binary-templates have higher SNR than the largest among $10^5$ DRW realizations (orange horizontal) none reach the SNR at injected values (marked with blue horizontal). We also note that there is a preference for higher eccentricities, which we attribute to a numerical artifact of the B24 hydrodynamical simulations. We discuss this further in \S~\ref{sec:discussion}.

We also find constraints on the binary inclination, with 86\% of these 536 high-SNR templates having values above greater than $85^\circ.$ However, the highest SNRs from this do not reach the true injected value, and we conclude that the initial search of sampling 10,000 binary-templates from a relatively sparse prior (Table~\ref{table:fiducial_binary_priors}) is insufficient to converge on the correct model because of the steep SNR dependence on $T_{\rm orb}$. 

\subsubsection{Step \#2: excised SLFs}

\begin{figure}
    \centering
    \includegraphics[width=\linewidth]{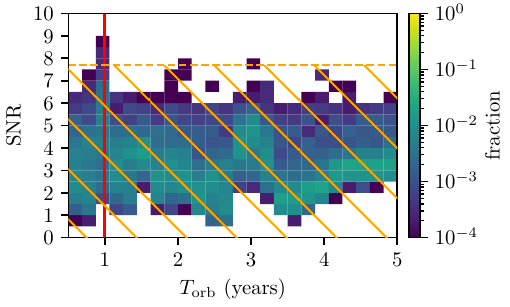}
    \caption{Step \#2, 2D histogram of 10,000 binary-templates with parameters sampled from a wide prior, binned by their binary parameters (horizontal axis) and matched filter SNRs (vertical axis). The color of each (parameter, SNR) bin indicates the fraction of templates that fall into that bin. The horizontal orange dashed line indicates the maximum SNR from $10^5$ DRW realizations. Injected orbital period is marked with a vertical red line. We find 4 binary-templates with orbital period $\sim 1 {\rm yr}$ and 1 binary-template with orbital period $\sim 2 {\rm yr}$ which have SNRs higher than DRWs.}
    \label{fig:step2_histograms}
\end{figure}

A better constraint on periodicity is necessary as it is the parameter that leads to the largest changes in the SNR (see Fig.~\ref{fig:SNR dependences}). From Step \#1, although we were able to constrain the injected orbital period to 1, 1.5 or 2.0 years, we need a more robust constraint. We take advantage of a second source of periodicity besides the flares: the hydrodynamical variability in the baseline. In order to analyze the baseline periodicity, we clip the flares from the light-curve, using the approximation that a lensing event lasts for the time it takes the lens to cross 4 Einstein radii of the source at alignment:
\begin{equation}\label{eq:lensing duration}
    t_{\rm lensing}=2\times \left(\frac{2R_{\rm E, source}}{v_{\rm orb}}\right) \approx 2\times \left(\frac{R_{\rm E, source}}{\pi a}T_{\rm orb}\right),
\end{equation} where $R_{\rm E,source}$ is the Einstein radius of the source and we assume a circular orbit to approximate the orbital velocity $v_{\rm orb}$. Then we remove the fraction $t_{\rm lensing}/T_{\rm orb}=2R_{\rm E, source}/\pi a$ of the brightest points from the light-curves, assuming that these correspond to the flares.

While Fig.~\ref{fig:step1_histograms} shows that $T_{\rm orb}=0.5, 1.0$ year has the highest SNR we purposely use $\tilde{T}_{\rm orb}=2.0$ yr as an incorrect Ansatz to calculate the semi-major axis $a$ in Eq.\ref{eq:lensing duration}. As shown in Fig.~\ref{fig:fiducial light curve}, this threshold still effectively deletes most of the flares. 

We then repeat the analysis of Step \#1, matching the same 10,000 templates to the mock light-curve with flares deleted. We present the SNR, orbital period distribution results of Step \#2 in Fig.~\ref{fig:step2_histograms}. We find that this is effective in identifying the injected periodicity, with 4 binary-templates at $T_{\rm orb}\sim1.00$ year and 1 binary-template at $T_{\rm orb}\sim2.00$ year having SNRs greater than SNRs from DRW realizations (these correspond to the 5 binary-templates in Fig.~\ref{fig:step2_histograms} which have SNR greater than the orange dashed line representing the maximum SNR from DRWs). Thus we can deduce, together from the information from Fig.~\ref{fig:step1_histograms} that there are periodically occurring flares from an edge-on $I\gtrsim85^\circ$ binary, and from the $T_{\rm orb}$ panel of Step \#2 that the true orbital period is $\sim 1$ yr. Using these new constraints, we set a narrowed prior of $T_{\rm orb}\in [0.95, 1.05]$ years around the orbital period and the binary inclination $I=[85^\circ,90^\circ]$, which are summarized in Table~\ref{table: narrow_prior}. 

\begin{table}[h]
\centering
\begin{tabular}{@{}cccc@{}}
\toprule
Rank & SNR & $T_{\rm orb}$ (yr) & $\Delta_{\rm opt}$ (d) \\
\midrule
1 & 8.8 & 1.0042 & 38 \\
2 & 8.4 & 1.0019 & 4 \\
3 & 8.2 & 1.0005 & 0 \\
4 & 8.1 & 0.9976 & 76 \\
5 & 7.9 & 2.0009 & 5 \\
\bottomrule
\end{tabular}
\caption{
Top-ranked binary-templates from Step~\#2. These templates are also the only five binary-templates with SNRs greater than the maximum from DRWs (orange dashed line in Fig.~\ref{fig:step2_histograms}). First row shows the injected parameters for the fiducial binary and each row below lists the ranking (by SNR), SNR value, recovered orbital period, and the optimal template shift $\Delta_{\rm opt}$ which maximizes the SNR. 
}
\label{table:step2_bestfits}
\end{table}

\begin{table}[h]
\centering
\begin{tabular}{ccc}
\toprule
Parameter & Fiducial $\hat{\theta}$ & Narrowed Prior \\
\midrule
$T_{\rm orb}$  (years) & 1 & $U[0.95, 1.05]$ \\
$I(^\circ)$ & $88$ & $U[85,90]$ \\
\bottomrule
\end{tabular}
\caption{The narrowed prior for the fiducial binary inferred in Steps \#1, \#2. Priors of parameters not mentioned in this table are kept the same as in Table~\ref{table:fiducial_binary} and Table~\ref{table:fiducial_binary_priors}.}
\label{table: narrow_prior}
\end{table}

\subsubsection{Step \#3: fine grid}

\begin{figure*}
    \centering
    \includegraphics[width=\linewidth]{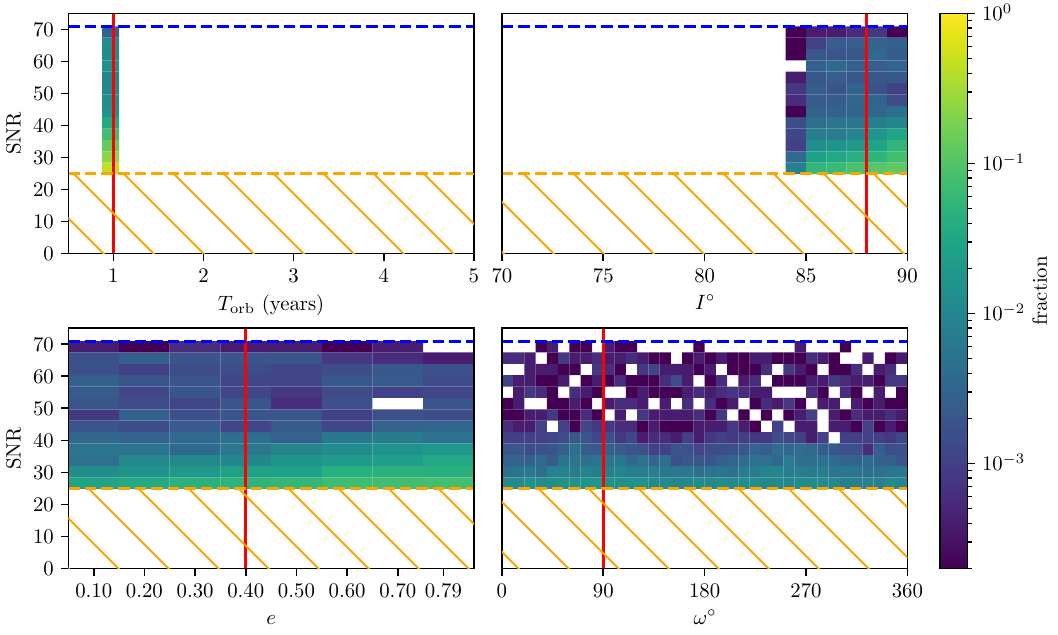}
    \caption{Step \#3, 2D histogram of 5006 out of 10,000 binary-templates with parameters sampled from a narrow prior, binned by their binary parameters (horizontal axis) and matched filter SNRs (vertical axis). The color of each (parameter, SNR) bin indicates the fraction of the selected 5006 binary templates that fall into that bin. These binary-templates have SNRs higher than the maximum SNR achieved with $10^5$ DRW realizations from Fig.~\ref{fig:null snrs}, marked with an orange dashed line. Injected binary parameters are marked with red lines. The SNR at injected values, marked by a blue dashed line, is now achieved as a result of our three-step procedure.}
    \label{fig:step3_histograms}
\end{figure*}

\begin{figure*}
    \centering
    \includegraphics[width=\linewidth]{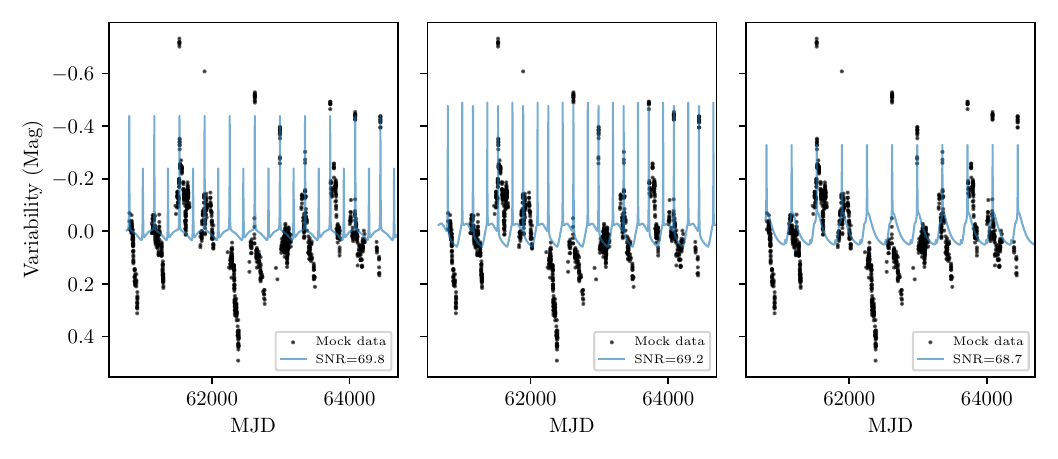}
    \caption{Final results (Step \#3): the three highest-SNR templates. The template SNRs are close to the SNR at the injected value 71 (see also Table~\ref{table:step3_bestfits}) and the templates match all of the self-lensing flares present in the mock data.}
    \label{fig:best_fit_templates}
\end{figure*}

\begin{figure*}
    \centering
    \includegraphics[width=1\linewidth]{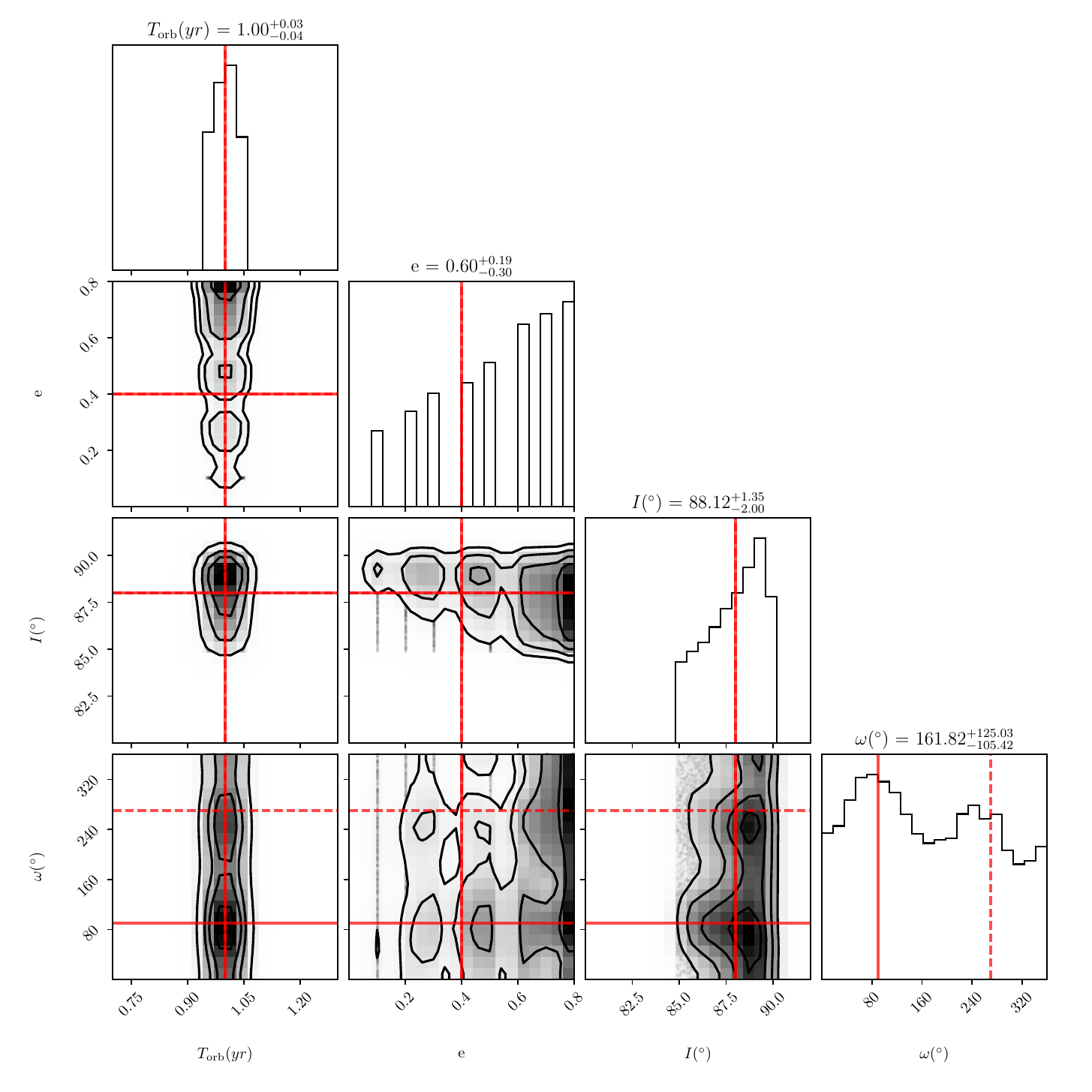}
    \caption{Corner plots of the results from Step \#3 with contours drawn using relative log-likelihood weights of $\ln\mathcal{L}-\ln\mathcal{L_{\rm max}}=-\frac{1}{2}({\rm \rho_{\rm max}}^2-{\rm \rho}^2)$. We recover the injected orbital period and binary inclination accurately. We find that the posterior distribution for the argument of periapsis also peaks at the injected orientation but we identify a bias towards higher eccentricities. See \S~\ref{sec:discussion} for further discussion.}
    \label{fig:step3_corner}
\end{figure*}

Finally, we Sobol-sample 10,000 binary-templates from the narrowed prior in $I, T_{\rm orb}$ of Table~\ref{table: narrow_prior}. Priors of parameters not mentioned in Table~\ref{table: narrow_prior} are kept the same as in Table~\ref{table:fiducial_binary} and Table~\ref{table:fiducial_binary_priors}. We present the parameter and SNR distribution of 5,006 binary-templates which exceed the DRW SNR threshold of 25 in Step \#1 in Fig.~\ref{fig:step3_histograms}. As a result we find that the SNRs of various binary-templates nearly reach the SNR at the injected values. We also plot the three highest-SNR binary-templates in Fig.~\ref{fig:best_fit_templates}, which now match all of the flares. The binary parameters of these binary-templates are listed in Table~\ref{table:step3_bestfits}. We also show the resulting parameter and SNR distribution in a corner plot (Fig.~\ref{fig:step3_corner}). The marginalized posterior contours are drawn using relative log-likelihood weights of $\ln\mathcal{L}-\ln\mathcal{L_{\rm max}}=-\frac{1}{2}({\rm \rho_{\rm max}}^2-{\rm \rho}^2)$, which follows from the log-likelihood $\ln \mathcal{L}(\theta)=\frac{1}{2}\rho^2(\theta)-\frac{1}{2}d^TC^{-1}d$ (assuming Gaussian-distributed noise) and the second term on the right hand side is independent of the binary parameters $\theta$. \cite{mfs_cosmology}. Here, $\rho_{\rm max}=69.8$ is the maximum SNR found among the 10,000 binary-templates. We recover the injected orbital period $T_{\rm orb}=1.00^{+0.03}_{-0.04}$ yr and the binary inclination $I=88.12^{+1.35^\circ}_{-2.00^\circ}$, where the values indicate the medians and the 16th, 84th percentile intervals of the marginalized posteriors of each parameter. We again find a preference for high eccentricities in Fig.~\ref{fig:step3_corner}. There are also two clear peaks at $\omega=90^\circ,270^\circ$ but the distribution is overall uniform, as can also be seen from the top-ranked binary-templates in Table~\ref{table:step3_bestfits}. This is because, for a fixed eccentricity, a change in the argument of periapsis primarily changes the timing of the flare within an orbit, but this change can be accounted for by the time shifts for each binary-template (see, for example, the $\omega$ and $\Delta_{\rm opt}$ values of the high-SNR binary-templates in Table~\ref{table:step3_bestfits}). Since the secondary flare was poorly sampled in this fiducial binary, it does not contribute to the SNR, and thus any value of $\omega$ can lead to a good fit with the mock light-curve as long as the other parameters are close to the injected parameters.  

While the best-fit values of $T_{\rm orb}$, $I$, and $\omega$ in Table~\ref{table:step3_bestfits} are consistent with the posterior shown in Fig.~\ref{fig:step3_corner}, the eccentricity behaves differently. 
Although the injected eccentricity $e=0.4$ yields the highest SNR at its peak (see Fig.~\ref{fig:SNR dependences}), the peak is extremely narrow in $(I,\omega)$ space. 
In contrast, templates with higher eccentricities ($e=0.7,0.79$) produce slightly lower SNRs but over a much broader region of parameter space. 
When marginalizing over $(I,\omega)$, this broader "plateau" contributes a larger total likelihood volume than the narrow peak at $e=0.4$, leading to an apparent preference for higher eccentricity in the corner plot. A visualization of this "plateau" in $(I,\omega)$ space is provided in the Appendix~\ref{sec:appendix}.

\begin{table}[h]
\centering
\begin{tabular}{@{}ccccccc@{}}
\toprule
Rank & SNR & $T_{\rm orb}$ (yr) & $I(^\circ)$ & $e$ & $\omega(^\circ)$ & $\Delta_{\rm opt}$ (d) \\
\midrule
Injected & 70.6 & 1.00 & 88.00 & 0.40 & 90.00  & 0  \\
\midrule
1 & 69.8 & 1.00 & 88.97 & 0.20 & 45.81  & 37   \\
2 & 69.2 & 1.00 & 89.48 & 0.20 & 304.05  & 133  \\
3 & 68.7 & 1.00 & 87.05 & 0.50 & 42.03  & 15 \\
\bottomrule
\end{tabular}
\caption{
Top-ranked binary-templates from Step~\#3 (Fig.~\ref{fig:best_fit_templates}). 
Like in Table~\ref{table:step1_bestfits}, the first row shows the injected parameters for the fiducial binary and each row below lists the ranking (by SNR), SNR value, corresponding binary parameters, and the optimal template shift $\Delta_{\rm opt}$ which maximizes the SNR.
}
\label{table:step3_bestfits}
\end{table}

\section{Discussion}
\label{sec:discussion}

\subsection{Limitations of mock binary light-curves}

\begin{figure}
    \centering
    \includegraphics[width=\linewidth]{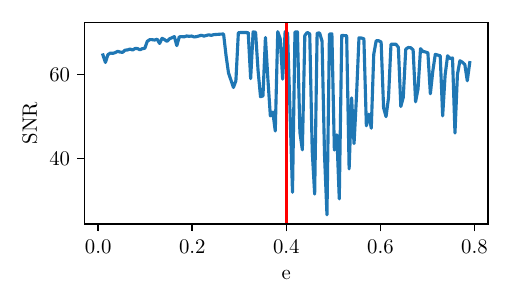}
    \caption{Matched filter SNR for the fiducial binary for a continuous prior in eccentricity. The oscillatory trend as a function of $e$ is artificial and arises from how \texttt{binline} samples the accretion switching between the two BH components. See \S~\ref{sec:discussion} for further discussion.}
    \label{fig:oscillating_e}
\end{figure}

Our analysis and results in \S~\ref{sec:Results} are based on the toy models for periodic binary light-curves in \texttt{binlite}.   
While this is suitable a first approximation, it leads to several caveats and limitations. 

The \texttt{binlite} light-curves are toy models which are built from blackbody emission from minidisks with equilibrium temperature profiles $T(r)\propto (M\dot{M}/r^3)^{1/4}$ and accretion rates measured in equal-mass, 2D Newtonian, hydrodynamical simulations with a locally isothermal equation of state. In reality, for quasars, the connection between accretion rates and luminosity is more complicated and not deterministic, so that light-curves should be directly extracted from more realistic simulations. For the construction of such light-curves, it will be necessary to incorporate the effects of radiation pressure, magnetic fields, general relativity, and the self-gravity of the gas into three-dimensional simulations.
\texttt{binlite} also needs to be extended to include  hydrodynamical simulations of unequal-mass binaries \cite{Siwek_2023} -- note that Paper IV and \cite{Kelley_2021} both find that mass ratios down to $q\sim 0.1$ will be relevant for self-lensing.   

We also had to use a discrete and sparse prior in $e$, because of a more subtle issue with \texttt{binlite} that results in a rapid oscillation of accretion rate ratio $Q\equiv\langle \dot{M}_2(t)\rangle/\langle\dot{M}_1(t)\rangle$ as a function of $e$. This is an artifact from the \texttt{binlite} accretion rates, which are generated from a single hydrodynamical simulation sweep in which the eccentricity is varied continuously from $e=0$ to $e=0.8$. Each eccentricity value therefore corresponds to a particular time in their sweep, from which \texttt{binlite} extracts its accretion rates. Along this sweep in eccentricity, the accretion rates $\dot{M}_1, \dot{M}_2$ onto the two minidisks fluctuate, causing the ratio $\dot{M}_1/ \dot{M}_2$ to fluctuate between values above and below unity as well. We demonstrate the accretion rate switching in Fig.~\ref{fig:oscillating_e} (and in Appendix~\ref{sec:appendix}), where "neighboring" eccentricities produce different flare profiles and SNRs due to frequent switches in which minidisk is brighter. This leads to a matched filter SNR profile that oscillates as a function of eccentricity whenever the accretion switching occurs.

As a result, lensing flare profiles look different as well, especially when the source and lens roles are reversed (for example, $\omega=90^\circ, 270^\circ$). This introduces a dependency on the argument of periapsis. For example, for certain eccentricities (e.g. $e=0.79$), the accretion ratio is closer to unity. Then, the flare profiles and the fit to the mock light-curve of the fiducial binary are similar when lens/source roles are swapped. Meanwhile for $e=0.4,$ the accretion ratio is unequal, so when lens/source roles are switched ($\omega=90^\circ\to270^\circ$) this leads to a much smaller flare because the dimmer disk becomes the source and the brighter disk becomes the lens. When marginalizing over $\omega$ in the corner plot (Fig.~\ref{fig:step3_corner}), we find that this causes an apparent bias towards higher eccentricities. We visualize this effect further in Appendix~\ref{sec:appendix}.

B24 also use the point-source approximation for self-lensing. In reality, most self-lensing binaries in LSST are expected to have masses of $\log(M/{\rm M_\odot})=7-9,$ mass ratios of $ q=0.1-1,$ and orbital periods of a few years \cite{Kelley_2021, park_2025}. According to \cite{dorazio2018}, these binaries will have accretion disc sizes that are comparable to the Einstein radius of the lenses, making the point-source approximation invalid. When the source is treated as a resolved disk and not a point, the effect is that the "Paczynski curve" (magnification vs time) is smeared out (dimmed and widened). However, Paper IV shows that the widening of the flares is more important than the dimming for detection rates. Thus incorporating the finite-source calculation of self-lensing in \texttt{binlite} is crucial as well and we expect that it could improve our results by making the flares less narrow and harder to miss. Beyond \texttt{binlite}, a future matched filter approach should also incorporate uncertainties on the binary total mass, Eddington ratio and redshift, which were assumed to be known. Developing mock binary light-curves incorporating various uncertainties could potentially be addressed by the concept of "fuzzy templates" \cite{fuzzy_templates}.

Finally, our analysis here was applied to a single band ($r$-band, chosen arbitrarily), whereas LSST will have data in 6 optical bands. Since SLFs are roughly achromatic between optical bands even in the finite-source limit, data from multiple bands will improve detections by increasing the number of data points and also the confidence of a detection. For these multi-wavelength matched filters (MMFs), correlations of quasar noise between optical bands will be required. 

\subsection{Different DRW realizations}
\begin{figure}
    \centering
    \includegraphics[width=\linewidth]{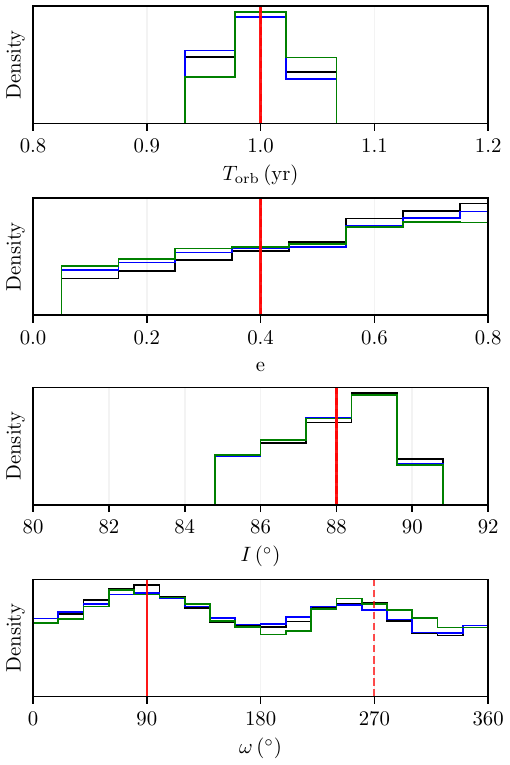}
    \caption{Distribution of 1D posteriors when applying Steps \#1-3 on the fiducial binary but with two other noise realizations. Black distributions are identical to the 1D distributions on the diagonal of the corner plot Fig.~\ref{fig:step3_corner} and blue, green distributions correspond to the two other noise realizations.}
    \label{fig:combined_posteriors}
\end{figure}

Another issue is that our analysis above was applied to a single mock binary, with a single noise realization.
To assess the robustness of our matched filter methods to the noise realization, we repeated Steps \#1, \#2, \#3 on the same fiducial binary parameters but for two more different realizations of the quasar DRW variability. Figure \ref{fig:combined_posteriors} shows the resulting 1D marginalized posterior distributions for the four inferred parameters. The black distributions are identical to the posteriors obtained in Fig.~\ref{fig:step3_corner}, while the blue and green curves show the distributions for the two additional DRW realizations. The three sets of distributions overlap closely, showing that our procedure is not sensitive to the specific DRW realization.

\subsection{Beyond the fiducial binary}

We next discuss the application of our matched filter procedure to binaries with parameters different from those of the fiducial binary. Given that flares dominate the matched filter SNR, the performance of our method depends sensitively on the sampling of the self-lensing flares. Binaries with longer-duration flares will have more data points covering each individual flare, which increases their detectability in our pipeline and improves periodicity recovery. Conversely, if some flares are missed due to observational gaps of the light-curve or because of a short flare durations, the detectability will degrade

Our method also partially relies (in Step \#2) on the hydrodynamical and Doppler variability in the baseline of the light-curve. In the case that the flares are poorly sampled, and quasar DRW noise dominates over baseline variability (e.g. small $A_{\rm hydro}/A_{\rm DRW}$ or $A_{\rm DB}/A_{\rm DRW}$), it will be difficult to recover the periodicity in Step \#2 and sample binary-templates in the vicinity of the injected parameters in Step \#3. However, we also found that Step \#1 identified multiple (though not all) flares in the mock light-curve (see Fig.~\ref{fig:optimal_shift}) with SNRs above the maximum of $10^5$ DRWs, despite sampling binary-templates from a wide prior (Table~\ref{table:fiducial_binary_priors}). This indicates that our pipeline can still flag such noisy light-curves, although the subsequent parameter recovery is less accurate when only a subset of flares is detected.

In our fiducial binary's mock light-curve (Fig.~\ref{fig:fiducial light curve}), the secondary self-lensing flares were much smaller compared to the primary and were also often missed due to sampling. In self-lensing binaries with brighter secondary flares that are adequately sampled and not drowned out by noise, we expect improved parameter recovery in $(e,\omega)$. For example, for fixed $\omega, T_{\rm orb},I$, a higher eccentricity leads to a shorter time interval between the flares followed by a longer waiting time for the next pair of flares to occur. On the other hand, for fixed $e, T_{\rm orb},I$, flare widths can also be used as an additional constraint. For example, the width of a self-lensing flare is shorter if $\omega\sim90^\circ$ when the secondary is at periapsis and moving quickly at closest alignment. 

We further tested whether our weak eccentricity dependence (Fig.~\ref{fig:SNR dependences}, top-left panel) is specific to the fiducial orientation $\omega=90^\circ$, where the line of sight is parallel to the semi-major axis and the flares always occur at periapsis and apoapsis for any eccentricity. By repeating the analysis at a different argument of periapsis $\omega=40^\circ$, we found a similar weak dependence of SNR on eccentricity. Again, this insensitivity arises because only one of the two flares per orbit contributes significantly to the SNR. Investigating the detectability and parameter estimation of binaries with a pair of bright and well-sampled self-lensing flares per orbit will be the subject of future work.

\subsection{Computational cost}

We consider computational costs necessary to search the 10-year catalog of LSST, which is projected to contain $N_{\rm QSO}=20-100$ million quasars \cite{Xin_Haiman_2021}. The number of CPU hours required for such a search will come from the amount of time to prepare the binary-templates $N_{\rm templates},$ which can be shared for the analysis of all quasar light-curves, and the amount of time to compute matched filter SNRs $N_{\rm QSO}\times N_{\rm templates}$ times. Although we plan to include other parameters like the mass ratio $q$ or the total mass $M$ of the binary in future searches, we assume that $N_{\rm templates}=10^4$ Sobol-sampled points will be more than sufficient to explore our parameter space.

On average, we estimate that preparing 1 \texttt{binlite} binary-template requires $t_{\rm pre}\sim0.4 \;{\rm seconds}$ and calculating 1 matched filter SNR requires $t_{\rm SNR}\sim0.2 \;{\rm seconds}$. Then the total CPU time required to search $N_{\rm QSO}$ light-curves is 

\begin{equation}
    \begin{split}
        t_{\rm CPU}=N_{\rm templates}\times t_{\rm pre} + N_{\rm QSO}\times N_{\rm templates}\times t_{\rm SNR}\\
        \simeq 22 \;{\rm million} \left(\frac{N_{\rm QSO}}{2\times10^7}\right) \left(\frac{N_{\rm templates}}{10^4}\right) \;{\rm CPU\; hours},
    \end{split}
\end{equation} where the template-preparation time (first term) is negligible compared to the total SNR calculation time (second term). $t_{\rm SNR}$ consists of the time to perform interpolation and dot-product operations using \texttt{scipy}~\cite{2020SciPy} and we plan to optimize these calculations further, which would proportionally reduce the computation costs. 

\section{Summary and conclusions}
\label{sec:conclusion}

It is estimated that there will be up to thousands of edge-on binaries exhibiting self-lensing flares lasting 30-days or longer and having orbital periods of 5 years or less (at least 2 orbits in 10 years). Using toy models of SMBHB photometric variability, we tested the efficacy of two statistical tools, LS periodograms and matched filters, to recover injected self-lensing flares in noisy light-curves. The conclusions of our study can be summarized as follows: 
\begin{itemize}
  \item None of the self-lensing flares have LS periodogram peaks exceeding the height of flares among 10,000 DRWs, even in ideal scenarios. A standard LS periodogram is unable to effectively recover periodicity from self-lensing flares.
  \item The matched filter SNR effectively separates self-lensing flares from DRW noise, yielding SNRs much greater than the maximum value found in $10^5$ DRW realizations.
  \item For our fiducial binary, a three-step matched filter procedure can identify periodicities as well as recover the injected periodicity and binary inclination.
  \item We find that constraining eccentricity is difficult because the secondary flare was not adequately sampled in our mock light-curve, leading to degeneracies among $e$, $\omega$, and an overall phase shift in the light-curve. We expect that bright and well-sampled secondary flares can break these degeneracies and help constrain $(e,\omega).$ 
\end{itemize}

We demonstrate that matched filters can potentially find self-lensing flares in LSST with reasonable computational cost. In the future, this approach will be refined by widening our prior to include the mass ratio as a free parameter, and more realistic light-curves that include 3D, general-relativistic effects, etc. Moreover, our single-band analysis should be extended to simultaneously analyzing all 6 bands of LSST. 

\acknowledgments
We acknowledge support by NASA grants 80NSSC22K0822 and 80NSSC24K0440 (ZH). 
VAV acknowledges support through the David and Lucile Packard Foundation, National Science Foundation under PHY-2019786 (The NSF AI Institute for Artificial Intelligence and Fundamental Interactions, http://iaifi.org/), AST-2433718, AST-2407922 and AST-2406110, as well as an Aramont Fellowship for Emerging Science Research. 
JD acknowledges support by a joint Columbia/Flatiron Postdoctoral Fellowship. Research at the Flatiron Institute is supported by the Simons Foundation. This research was supported in part by the National Science Foundation under Grant No. NSF PHY-1748958 to the Kavli Institute for Theoretical Physics (KITP). This research has made use of NASA's Astrophysics Data System. The authors acknowledge the Texas Advanced Computing Center (TACC) at The University of Texas at Austin for providing computational resources.

{\it Software:} {\tt python} \citep{travis2007,jarrod2011}, {\tt numpy} \citep{walt2011}, {\tt matplotlib} \citep{hunter2007}, {\tt scipy}\cite{2020SciPy}, {\tt astroML}\cite{astroMLText}, {\tt binlite}\cite{D’Orazio_2024}, {\tt corner} \cite{corner}

\section*{Data Availability}
The data underlying this article will be shared on reasonable request to the corresponding author.

\appendix
\section{SNR Surfaces} \label{sec:appendix}
\begin{figure*}
    \centering
    \includegraphics[width=\linewidth]{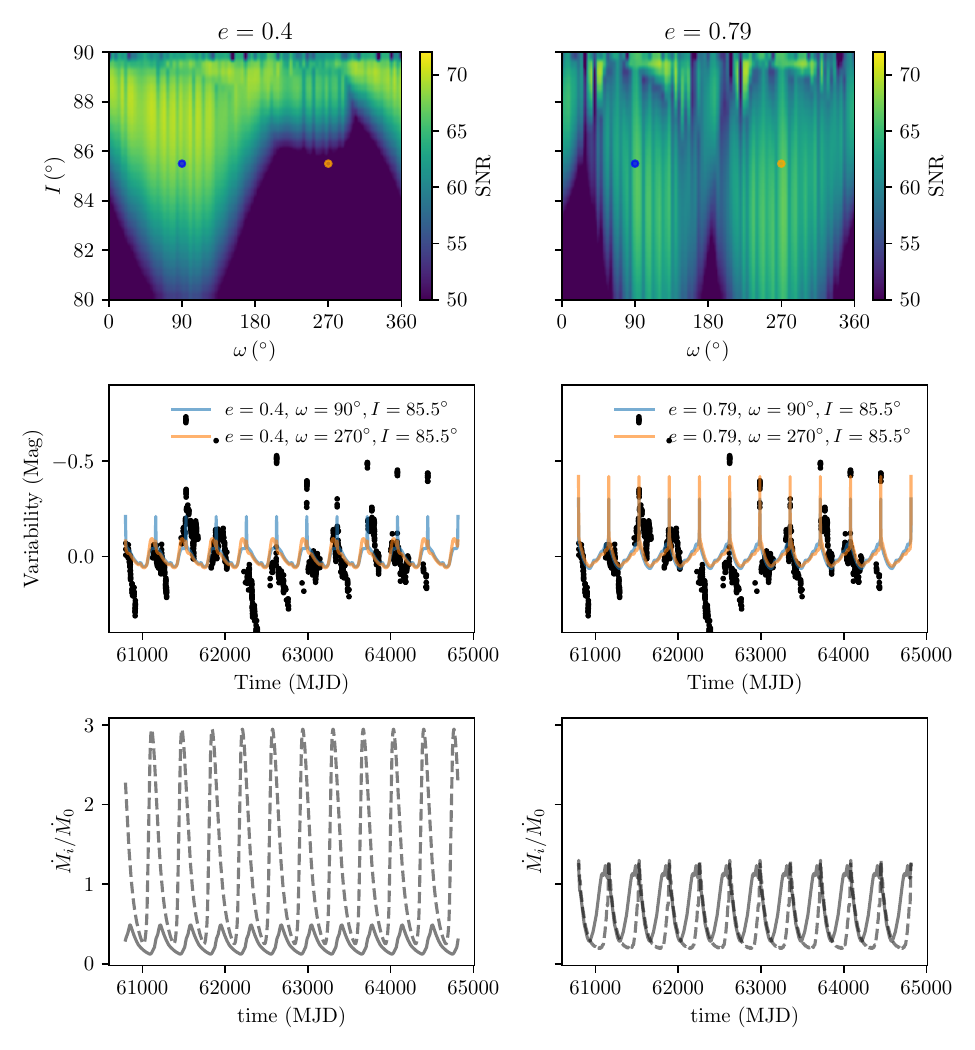}
    \caption{\textit{First row:} SNR surfaces for $e=0.4$ (injected value) and $e=0.79$ of the fiducial binary in $(I,\omega)$ space, where the brightest color indicates the SNR at the injected values (71). The two blue, orange points on each panel indicate points that characterize the $e$ "plateau" described in \S~\ref{sec:Results}. For $e=0.4,$ at the blue point $(\omega,I)=(90^\circ, 85.5^\circ),$ the SNR is 67.9, whereas at the orange point $(\omega,I)=(270^\circ, 85.5^\circ),$ the SNR is 46.2. However for $e=0.79,$ at the blue point $(\omega,I)=(90^\circ, 85.5^\circ),$ the SNR is 65.5, whereas at the orange point $(\omega,I)=(270^\circ, 85.5^\circ),$ the SNR is 65.2. There is a clear difference in SNR for $e=0.4$ vs $e=0.79$ when the lensing flare is viewed from the opposite side. \textit{Second row:} Binary-template light-curves generated from the blue and orange points in the $(I,\omega)$ parameter space in the first row. There is a large difference in the flare heights for $e=0.4$ for $\omega=90^\circ,270^\circ$ whereas for $e=0.79$ the difference is much smaller. \textit{Third row:} The accretion rates of the primary (dashed) and secondary (solid) minidisks, determined by the \texttt{binlite} simulation sweep. These are plotted in gray to emphasize that they depend only on eccentricity and not on $(I,\omega)$. The accretion-rate contrast between the two minidisks is large for $e=0.4$ and much smaller for $e=0.79$, explaining the asymmetric versus symmetric flare behavior seen in the first two rows.}
    \label{fig:snr_surfaces}
\end{figure*}


\bibliography{bibfile}

\appendix

\end{document}